\documentclass[
prm,
twocolumn,
superscriptaddress,
 amsmath,amssymb,
 aps,
longbibliography,
]{revtex4-2}
\usepackage{lmodern}
\usepackage[utf8]{inputenc}
\usepackage[T1]{fontenc}
\usepackage{xcolor}
\usepackage{babel}
\usepackage{array}
\usepackage{booktabs}
\usepackage{textcomp}
\usepackage{amsmath}
\usepackage{amstext}
\usepackage{graphicx}
\usepackage{subscript}
\usepackage{comment}
\usepackage{bm}
\usepackage{setspace}
\usepackage{soul}

\usepackage[unicode=true,pdfusetitle,
 bookmarks=true,bookmarksnumbered=false,bookmarksopen=false,
 breaklinks=false,pdfborder={0 0 1},backref=false,colorlinks=true]
 {hyperref}
\hypersetup{
 linkcolor=magenta, citecolor=blue, urlcolor=blue, pdfstartview=XYZ, plainpages=false, pdfpagelabels}
\DeclareMathAlphabet{\mathsfit}{\encodingdefault}{\sfdefault}{m}{sl}
\SetMathAlphabet{\mathsfit}{bold}{\encodingdefault}{\sfdefault}{bx}{sl}
\newcommand{\tens}[1]{\bm{\mathsfit{#1}}}
\newcommand{\tenscomp}[1]{\mathsfit{#1}}
\newcommand{\WidthFigure}{\columnwidth}

\makeatletter
\PassOptionsToPackage{square,comma}{natbib}
\RequirePackage{natbib}
\usepackage{float}
\floatstyle{plaintop}
\restylefloat{table}
\newcommand{\lyxmathsym}[1]{\ifmmode\begingroup\def\b@ld{bold}
  \text{\ifx\math@version\b@ld\bfseries\fi#1}\endgroup\else#1\fi}

\providecommand{\tabularnewline}{\\}
\makeatother
\begin{document}



\title{Wigner heat transport in LaPO$_{4}$-based alloys for thermal barrier coatings}
\title{Dual wave-particle heat transport in LaPO$_{4}$-based alloys for thermal barrier coatings}
\title{First-principles characterization of thermal conductivity in LaPO$_{4}$-based alloys}
\newcommand*\mycommand[1]{\texttt{\emph{#1}}}
\author{Anees Pazhedath}
\affiliation{U Bremen Excellence Chair, Bremen Center for Computational Materials Science, and MAPEX Center for Materials and Processes, University of Bremen, D-28359 Bremen, Germany}
\author{Lorenzo Bastonero}
\affiliation{U Bremen Excellence Chair, Bremen Center for Computational Materials Science, and MAPEX Center for Materials and Processes, University of Bremen, D-28359 Bremen, Germany}
\author{Nicola Marzari}
\affiliation{U Bremen Excellence Chair, Bremen Center for Computational Materials Science, and MAPEX Center for Materials and Processes, University of Bremen, D-28359 Bremen, Germany}
\affiliation{Theory and Simulation of Materials (THEOS) and National Centre for Computational Design and Discovery of Novel Materials (MARVEL), \'Ecole Polytechnique F\'ed\'erale de Lausanne, Switzerland}
\affiliation{Laboratory for Materials Simulations, Paul Scherrer Institute, 5232 Villigen PSI, Switzerland}
\author{Michele Simoncelli}
\email{ms2855@cam.ac.uk}
\affiliation{Cavendish Laboratory, Theory of Condensed Matter Group, University of Cambridge, United Kingdom} 

\begin{abstract}
Alloys based on lanthanum phosphate (LaPO$_{4}$) are often employed as thermal barrier coatings, due to their low thermal conductivity and structural stability over a wide temperature range. 
To enhance the thermal-insulation performance of these alloys, it is essential to comprehensively understand the fundamental physics governing their heat conduction.
Here, we employ the Wigner formulation of thermal transport in conjunction with first-principles calculations to elucidate how the interplay between anharmonicity and compositional disorder determines the thermal properties of La$_{1{-}x}$Gd$_{x}$PO$_{4}$ alloys, and discuss the fundamental physics underlying the emergence and coexistence of particle-like and wave-like heat-transport mechanisms. 
We also show how the Wigner transport equation describes correctly the thermodynamic limit of a compositionally disordered crystal, while the Boltzmann transport equation does not.
Our predictions for microscopic vibrational properties (temperature-dependent Raman spectrum) and for macroscopic thermal conductivity are validated against experiments. Finally, we leverage these findings to devise strategies to optimize the performance of thermal barrier coatings.
\end{abstract}
\maketitle
\section{INTRODUCTION}

Improving the thrust and efficiency of airbreathing jet engines requires increasing the temperature of the gas at the entry of their turbine cascade (inlet temperature)  \cite{padture_advanced_2016}. Since the 1970s, significant research efforts have been made to develop materials capable of operating at increasingly higher temperatures, and current turbines employ superalloys blades covered by thermal barrier coatings (TBCs) \cite{clarke_thermal-barrier_2012,zhang_al2o3-modified_2020,tejero-martin_review_2021,leng_solution_2022}.
TBCs are critical in determining the performance and lifespan of turbines: they protect the blades from thermal stresses, allowing operational temperatures higher than the melting point of the superalloy.
Thus, one of the key objectives  \cite{miller_thermal_1997,liu_advances_2019,xu_evolution_2015} in current research on TBCs is to find materials with the lowest possible thermal conductivity  \cite{clarke_thermal-barrier_2012,padture_advanced_2016}.

Hitherto, most of the progress on improving the thermal-insulation performance of TBC materials has relied on experiments and trial-and-error efforts, which hinted that the presence of compositional disordered in highly anharmonic materials can be beneficial for their thermal-insulation performance \cite{wan_glass-like_2010,wang_role_2014}.
However, a first-principles understanding of the microscopic physics governing heat conduction in these materials is missing, preventing their systematic optimization. 
The absence of such understanding on how the interplay between anharmonicity and compositional disorder affects the transport properties of TBC can be traced back to limitations of established first-principles approaches for thermal transport in solids, namely first-principles molecular dynamics (FPMD)  \cite{Martin2022,bouzid_thermal_2017,duong_thermal_2019,marcolongo2016microscopic,carbogno2017ab,puligheddu2017first} and the linearized Boltzmann transport equation for phonons (LBTE)  \cite{broido_intrinsic_2007,PhysRevB.88.045430,paulatto_first-principles_2015,cepellotti_phoebe_2022,togo_first-principles_2023,chaput_direct_2013,tadano_anharmonic_2014,carrete_almabte_2017,lindsay_perspective_2019}. Specifically, despite recent advances  \cite{knoop_ab_2023,knoop_anharmonicity_2023}, FPMD approaches still have a high computational cost, which practically limits their application to materials having nanometric disorder length scale (simulations cells containing a few hundred atoms).
On the other hand, the LBTE accounts exclusively for particle-like (intraband) heat transport mechanisms and misses wave-like interband tunneling; thus, it is in principle accurate only in weakly anharmonic crystals characterized by well-separated phonon bands \cite{simoncelli_unified_2019}.
The LBTE with a perturbative description of compositional (mass) disorder  \cite{PhysRevB.27.858} has been shown to successfully describe the thermal properties of SiGe alloys \cite{PhysRevLett.106.045901}, \textit{i.e.} weakly anharmonic materials in which low-frequency vibrational modes dominate transport  \cite{thebaud_success_2020}. However, such a perturbative scheme fails in strongly anharmonic systems where transport is not dominated by low-frequency modes  \cite{thebaud_perturbation_2022}.
Thermal insulators and alloys for TBCs  \cite{Wigner_paper,luo_vibrational_2020} belong to this class; thus, the effect of compositional disorder on the thermal conductivity of these materials cannot be described using the perturbative treatment within the LBTE.

These limitations can be overcome relying on the recently introduced Wigner Transport Equation (WTE) \cite{simoncelli_unified_2019}. 
Specifically, the WTE generalizes the LBTE by accounting not only for the particle-like propagation of populations of phonons, but also for wave-like tunnelling between vibrational eigenstates with energy difference smaller than their linewidths.
The former transport mechanism is commonly called ``intraband'', or ``band-diagonal'' transport; the latter is ofter referred to as ``interband'' transport originating from the band-off-diagonal elements of the Wigner density matrix, also known as  ``coherences''\cite{Wigner_paper}.
Accounting for both these transport mechanisms allows to unify under the same formal equation the LBTE for weakly anharmonic crystals and the Allen-Feldman equation for disordered solids  \cite{PhysRevLett.62.645}, also covering all the intermediate cases---such as alloys for TBC materials---in which anharmonicity  \cite{Wigner_paper,di_lucente_crossover_2023,xia_microscopic_2020,jain_multichannel_2020,Kpal_microscopic_2021, caldarelli_many-body_2022} and disorder  \cite{isaeva2019modeling,lundgren_mode_2021,liu_unraveling_nodate,simoncelli_thermal_2023,harper_vibrational_2023} are both relevant. 

Here, we combine first-principles calculations with the Wigner formulation to unveil the microscopic heat conduction mechanisms in LaPO$_{4}$ \cite{liu_advances_2019,https://doi.org/10.1111/j.1551-2916.2009.03244.x,Dong_2019,ZHANG20188818,shijina_very_2016}, in its solid solutions with GdPO$_{4}$, and its composites with La$_2$Zr$_2$O$_7$. These materials are employed in TBCs  \cite{liu_advances_2019,https://doi.org/10.1111/j.1551-2916.2009.03244.x,Dong_2019, ZHANG20188818, shijina_very_2016} due to their low thermal conductivity \cite{https://doi.org/10.1111/j.1551-2916.2009.03244.x,Dong_2019,ZHANG20188818,shijina_very_2016}, high melting point  \cite{https://doi.org/10.1111/j.1151-2916.1987.tb04890.x}, chemical durability  \cite{RAFIUDDIN2018631}, structural stability  \cite{CLAVIER2011941} and large thermal expansion coefficient  \cite{MIN2001939} over a wide temperature range. 
After characterizing the properties of the heat carriers in LaPO$_4$, we show that microscopic particle-like and wave-like transport mechanisms emerge and coexist in this material, and we discuss how they contribute to the macroscopic thermal conductivity. Next, we analyze how atomic-scale compositional disorder affects thermal transport in La$_{1-x}$Gd$_{x}$PO$_{4}$ alloys, describing La-Gd mass-substitutional disorder explicitly using models containing up to 5184 atoms. Finally, we use the continuum Maxwell's model for the  effective conductivity of composites \cite{maxwell_treatise_1873,sevostianov_maxwells_2019} to discuss how the thermal conductivity is affected by the compositional disorder at the micrometer scale, 
investigating composites containing micrometer-sized grains of La$_{2}$Zr$_{2}$O$_{7}$ and LaPO$_4$ at different concentrations.

\section{Wigner formulation for thermal transport}
\label{sec:Wigner_formulation}

The Wigner transport equation (WTE)  \cite{simoncelli_unified_2019, Wigner_paper} describes thermal transport in solids accounting for the interplay between structural disorder, anharmonicity, and Bose-Einstein statistics of vibrations. Such an equation provides a comprehensive approach to thermal transport in solids, allowing to describe structurally ordered ``simple crystals'' with interband spacings much larger than the linewidths  \cite{PhysRevX.10.011019,dragasevic_viscous_2023}, glasses  \cite{liu_unraveling_nodate,simoncelli_thermal_2023,harper_vibrational_2023}, as well as the intermediate regime of ``complex crystals'' with interband spacings smaller than the linewidths  \cite{Wigner_paper,di_lucente_crossover_2023,xia_microscopic_2020,jain_multichannel_2020}. 
In the following, we summarize the salient features of the WTE. For complex crystals having ultralow thermal conductivity, the WTE can be accurately solved using the single-mode relaxation-time approximation (SMA)  \cite{PhysRevLett.106.045901,lindsay_first_2016,simoncelli_thermal_2023}, since in these cases the SMA yields results practically indistinguishable from the exact solution  \cite{simoncelli_unified_2019,Wigner_paper,simoncelli_thermal_2023,jain_multichannel_2020,xia_microscopic_2020}. Within the SMA, the Wigner conductivity expression assumes the following compact form:
\begin{equation}
\begin{split}
&\kappa^{\alpha\beta}{=}\frac{1}{\mathcal{V}{N_{\rm c}} }{\sum_{\bm{q},s,s'}}\!
\frac{\omega(\bm{q})_s{+}\omega(\bm{q})_{s'}}{4}\!\left(\frac{C(\bm{q})_{s}}{\omega(\bm{q})_{s}}{+}\frac{C(\bm{q})_{s'\!}}{\omega(\bm{q})_{s'\!}}\right)\!\tenscomp{v}^{\alpha}(\bm{q})_{s,s'}\!\\
&{\times}\tenscomp{v}^{\beta}(\bm{q})_{s',s}
\frac{\frac{1}{2}\big(\Gamma(\bm{q})_s{+}\Gamma(\bm{q})_{s'}\big)}{\big(\omega(\bm{q})_s{-}\omega(\bm{q})_{s'}\big)^2+\frac{1}{4}\big(\Gamma(\bm{q})_s{+}\Gamma(\bm{q})_{s'}\big)^2}
\;,
\label{eq:thermal_conductivity_combined}
\end{split}
\raisetag{5mm}
\end{equation}
where $\hbar\omega(\bm{q})_s$ is the energy of the phonon having wavevector $\bm{q}$ and mode index $s$, which carries the specific heat 
\begin{equation}
C(\bm{q})_s{=}C[\omega(\bm{q})_s]{=}\frac{\hbar^2\omega^2(\bm{q})_s }{k_{\rm B} T^2} \bar{\tenscomp{N}}(\bm{q})_s\big[\bar{\tenscomp{N}}(\bm{q})_s{+}1\big],  
\label{eq:quantum_specific_heat_A}
\end{equation}
where $\bar{\tenscomp{N}}(\bm{q})_s{=}[\exp(\hbar \omega(\bm{q})_s/k_{\rm B}T){-}1]^{-1}$ is the Bose-Einstein distribution at temperature $T$,
$\tenscomp{v}^{\alpha}(\bm{q})_{s,s'}$ is the velocity operator coupling eigenstates $s$ and $s'$ at the same wavevector $\bm{q}$ ($\alpha$ denotes a Cartesian direction)  \cite{Wigner_paper}, $N_{\rm c}$ is the number of $\bm{q}$-points entering in the $\bm{q}$-summation and $\mathcal{V}$ is the primitive-cell volume;
finally,
$\hbar\Gamma(\bm{q})_s=\hbar\Gamma(\bm{q})^{\rm anh}_s{+}\hbar\Gamma(\bm{q})^{\rm iso}_s{+}\hbar\Gamma(\bm{q})^{\rm bnd}_s$ is the total linewidth, determined by
anharmonicity  \cite{paulatto_anharmonic_2013,PhysRevB.88.045430} ($\hbar\Gamma(\bm{q})^{\rm anh}_s$), isotopic-mass disorder  \cite{PhysRevB.27.858} ($\hbar\Gamma(\bm{q})^{\rm iso}_s$) and grain-boundary scattering  \cite{fugallo_thermal_2014,casimir_note_1938} ($\hbar\Gamma(\bm{q})^{\rm bnd}_s$), see Appendix~\ref{sec:contributions_to_phonon_linewidths} for details.

In crystals, it is useful to resolve the WTE conductivity~(\ref{eq:thermal_conductivity_combined}) as the sum of two terms, $\kappa{=}\kappa_P{+}\kappa_C$  \cite{Wigner_paper}. The term $\kappa_P$ is referred to as ``populations conductivity''  \cite{simoncelli_unified_2019,Wigner_paper} and is determined by the diagonal ($s{=}s'$) or perfectly degenerate ($s{\neq}s'$ with $\omega(\bm{q})_{s}{=}\omega(\bm{q})_{s'}$) terms in the summation in expression~(\ref{eq:thermal_conductivity_combined}). Specifically, $\kappa_P$ can be written as  $\kappa_P^{\alpha\alpha}{=}\frac{1}{\mathcal{V} N_c }\sum_{\bm{q}s}C[\omega(\bm{q})_{s}]{\tenscomp{v}^{\alpha}(\bm{q})_{s,s}}\Lambda^\alpha(\bm{q})_s$; this expression shows that $\kappa_P$ describes a heat-transport mechanism in which heat carriers transport the energy $\hbar\omega(\bm{q})_{s}$ and propagate particle-like with velocity ${\tenscomp{v}^{\alpha}(\bm{q})_{s,s}}$ over the mean-free path $\Lambda^\alpha(\bm{q})_{s}{=}{\tenscomp{v}^{\alpha}(\bm{q})_{s,s}}[\Gamma(\bm{q})_s]^{-1}$, in analogy with particles in a classical gas.
In contrast, the non-degenerate off-diagonal elements (referred to as ``coherences''  \cite{simoncelli_unified_2019, Wigner_paper}) do not have an absolute energy but are characterized by an energy difference $\hbar\omega(\bm{q})_{s}{-}\hbar\omega(\bm{q})_{s'}$; they describe conduction through a wave-like tunneling mechanism between pairs of phonon bands (a mechanisms bearing analogies to the electronic Zener interband tunneling  \cite{krieger_quantum_1987}). In Eq.~(\ref{eq:thermal_conductivity_combined}), non-degenerate off-diagonal elements determine the coherences conductivity $\kappa_C$  \cite{simoncelli_unified_2019,Wigner_paper}. It has been shown in Refs.  \cite{simoncelli_unified_2019, Wigner_paper,simoncelli_temperature-invariant_2024} that in simple crystals particle-like mechanisms dominate over the wave-like tunnelling and thus $\kappa_{_{\rm P}}{\gg}\kappa_{_{\rm C}}$, while in complex crystals both these mechanisms co-exist and may have comparable strength, implying $\kappa_{_{\rm P}}{\sim}\kappa_{_{\rm C}}$. Finally, Refs.~\cite{simoncelli_thermal_2023,harper_vibrational_2023,fiorentino_hydrodynamic_2023} have shown that in strongly disordered oxide glasses $\kappa_{_{\rm P}}{\ll}\kappa_{_{\rm C}}$. 

\section{RESULTS AND DISCUSSION}\label{Results and discussions}
\subsection{LaPO$_{4}$: vibrational properties \& Raman spectra} 
In order to elucidate the microscopic physics that governs thermal transport in LaPO$_4$-based alloys, we start by computing from first principles the microscopic vibrational properties appearing in the thermal conductivity expression~(\ref{eq:thermal_conductivity_combined}) for the fundamental component LaPO$_4$.
Specifically, we employ density-functional theory to optimize the geometry (see Fig.~\ref{optimized-str} in Appendix \ref{sec: computational methods}), to compute the vibrational harmonic frequencies \cite{RevModPhys.73.515} (see Appendix \ref{sec: computational methods}, in particular the phonon dispersions are shown in Fig.~\ref{phonon-disp}, and the phonon density of states is shown in Fig.~\ref{Alloy-config-DOS}) and anharmonic third-order linewidths  \cite{togo_first-principles_2023,paulatto_first-principles_2015,PhysRevB.88.045430} (see Fig.~\ref{LW_fit-TC-compare} in Appendix~\ref{sec: computational methods}).
These quantities are calculated employing the standard perturbative approach, where frequencies and velocity operators are determined at the harmonic level and considered temperature-independent, and the anharmonic linewidths depend on temperature through the Bose-Einstein distribution (see Eq.~(\ref{eq:anh_linewidth}) in Appendix~\ref{sec:contributions_to_phonon_linewidths}).
\begin{figure}[b]
    \centering
    \includegraphics[width=\WidthFigure]{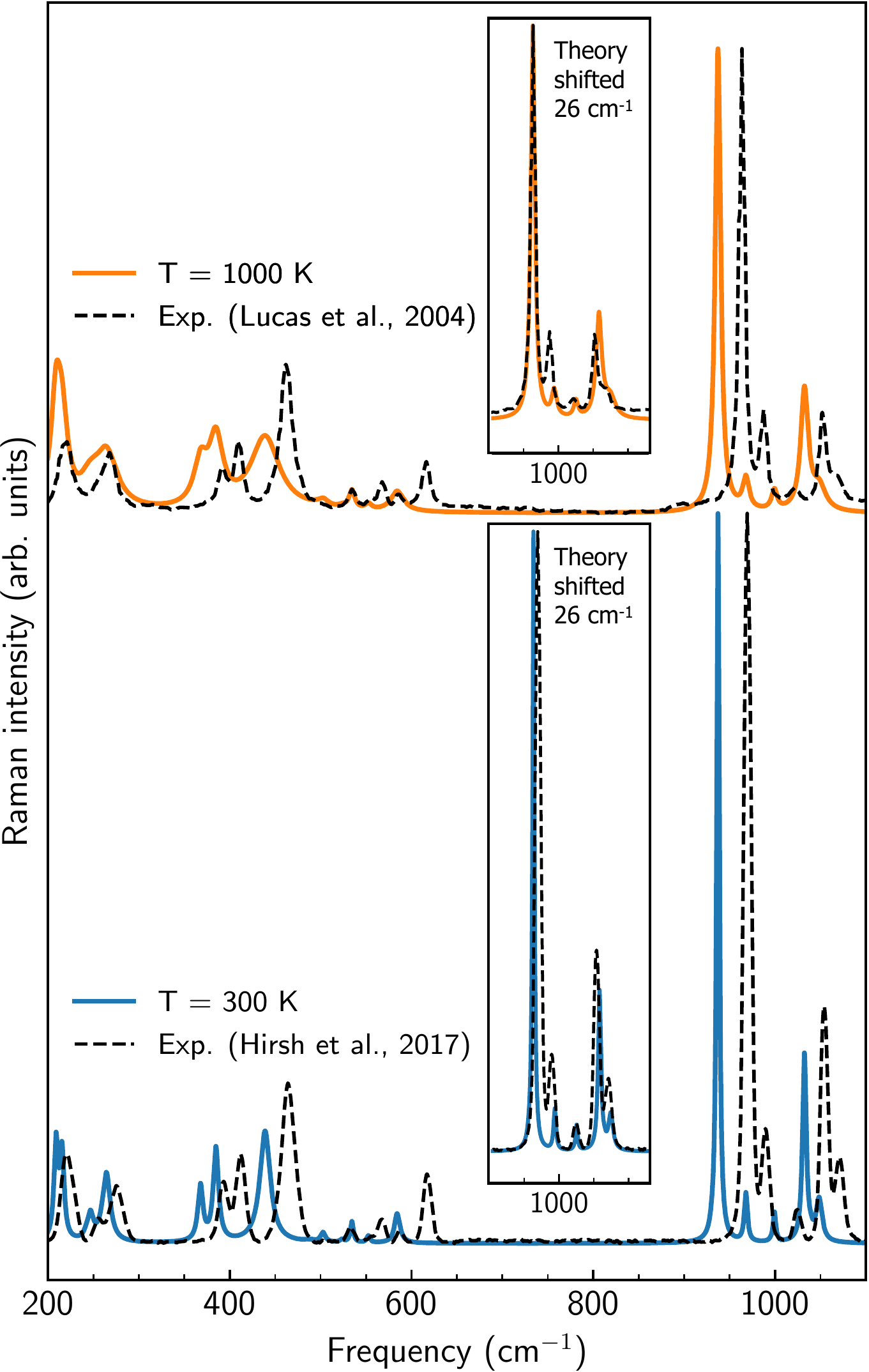}
    \caption{\textbf{Temperature-dependent Raman spectra.} Solid lines are theory at 1000 K (top) and 300 K (bottom); dashed lines are experimental data, from Lucas et al.  \cite{LUCAS20041312} at 1000 K (top) and from Hirsh et al.  \cite{HIRSCH201782} at 300 K (bottom). 
    In the insets, the theoretical spectra have been rigidly shifted by 26 cm$^{-1}$ to higher frequencies (right) to ease the comparison between the broadening of the theoretical and experimental spectra. }
  \label{fig:raman}
\end{figure}
In order to assess the accuracy of this perturbative approach in describing the actual frequencies and linewidths, we employ the theoretical frequencies and linewidths to predict the temperature dependence of the experimentally measurable non-resonant Raman spectra  \cite{cusco_temperature_2016}, 
\begin{equation}
    I(\omega, T) 
    \propto 
    \sum_{s}
    I_{s}(T) 
    \frac{ (\Gamma_s + \Gamma_{\rm ins})/2 }
    {(\omega -\omega_s)^2 + (\Gamma_s+\Gamma_{\rm ins})^2/4},
    \label{eq:raman}
\end{equation}
where $I_s(T)$ is the Raman intensity of the phonon mode $s$ computed within the Placzek approximation and using the powder average formula 
(Eq. (5) in Ref.~\cite{PhysRevB.71.214307}) including the laser-frequency factor at experimental conditions; $\omega_s$ and $\Gamma_s$ are the bare phonon frequencies and total linewidths at $\bm{q}=\mathbf{0}$. The linewidths are the full width at half maximum, related to the lifetime as $\tau_s=[\Gamma_s]^{-1}$, and are determined by both anharmonicity ($\Gamma^{\rm anh}_s$) and isotopic-mass disorder ($\Gamma^{\rm iso}_s$), \textit{i.e.} $\Gamma_s{=}\Gamma^{\rm iso}_s{+}\Gamma^{\rm anh}_s$, respectively. Finally, $\Gamma_{\rm ins} {=} 2$ cm$^{-1}$ accounts for the  instrumental broadening \cite{tuschel2020spectral} affecting the experiments \cite{HIRSCH201782,LUCAS20041312} with which we compare our calculations. In this approach the temperature dependence of the Raman spectra originates from the Bose-Einstein occupation numbers appearing in the intensity $I_s(T)$, and from the linewidths appearing in the Lorentzian broadening, as in previous work \cite{Wigner_paper}.  

We show in Fig.~\ref{fig:raman} a comparison between the theoretical and experimental Raman intensities in powder samples at 300 K (experiments by  \citet{HIRSCH201782}) and at 1000 K (experiments by  \citet{LUCAS20041312}).
Then, we note that the theoretical and experimental spectra are systematically shifted by approximately 26 cm$^{-1}$.  
Systematic, rigid shifts between the theoretical and experimental spectra
comparable to those observed here are common in the literature, and the small relative discrepancies ($\lesssim 2.5\%$ for the optical modes) observed here are within the accuracy with which DFT predictions usually describe experiments
\cite{lazzeri_first-principles_2003,bagheri_high-throughput_2023,liang_high-throughput_2019}. We also note that energy shifts in the symmetric or asymmetric PO$_4$ stretching modes around 1000 cm$^{-1}$ may be related to the presence of water in the experimental samples \cite{LUCAS20041312,HIRSCH201782,liang_high-throughput_2019}, which is not accounted for in our calculations.
More importantly, the positions of the experimental peaks are essentially unaffected by temperature; this indicates that in LaPO$_{4}$, the temperature renormalization of the vibrational frequencies, not accounted for in the standard perturbative treatment of anharmonicity employed here, is unimportant. 
In the insets of Fig.~\ref{fig:raman}, we highlight how the broadening of the experimental Raman peaks --- which is determined mainly by anharmonicity and is negligibly affected by compositional disorder \cite{HIRSCH201782} --- is in agreement with our theoretical predictions at all temperatures for the most intense, high-frequency part of the Raman spectrum. 
We recall that the anharmonic linewidths generally increase with frequency and temperature~ \cite{Wigner_paper,simoncelli_thermal_2023,di_lucente_crossover_2023,harper_vibrational_2023} (see  Fig.~\ref{LW_fit-TC-compare} in Appendix~\ref{sec:accounting_for_anharmonicity_at_a_reduced_computational_cost}).
Thus, from the good agreement between the theoretical and experimental broadening at high frequencies --- where anharmonic effects are usually largest --- we infer that the standard perturbative approach employed here provides a description of the anharmonic microscopic vibrational properties of LaPO$_4$ that is sufficiently accurate for our purposes.
Finally, we note that the low-frequency part of the experimental Raman spectrum at 1000 K is sharper than the theoretical predictions. Such behavior may be due to the occurrence of grain coalescence in the sample of Lucas et al.  \cite{LUCAS20041312}, which could cause partial crystallization and departure from our calculations that assume a powder sample. Additional details on the Raman simulations are reported in Appendix~\ref{sub:raman_spectrum}.

\subsection{Thermal conductivity of LaPO$_{4}$}\label{TC-primcell} 

\begin{figure}[t]
\includegraphics[width=\columnwidth]{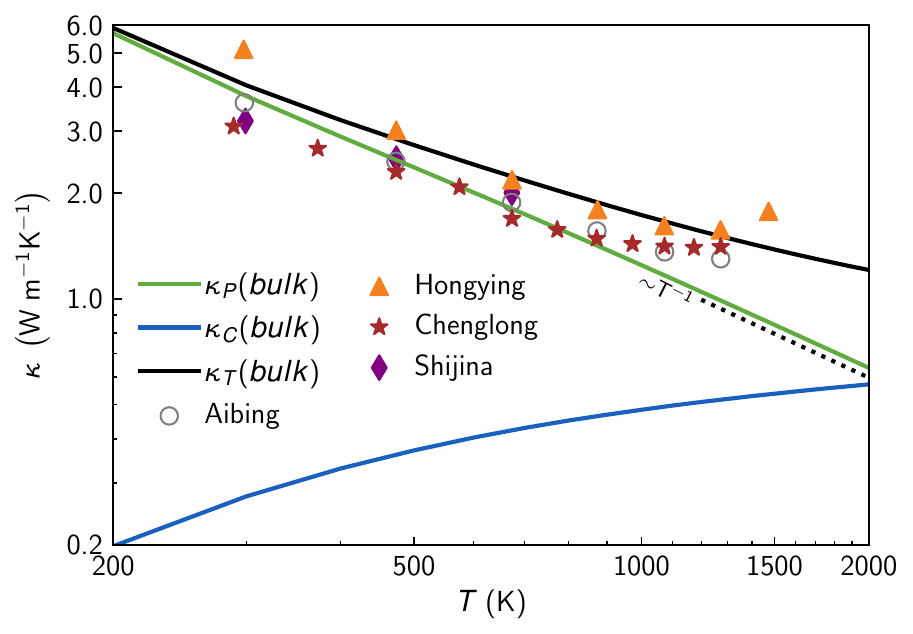}\caption{\textbf{Thermal conductivity of LaPO$_{4}$.} 
Green, populations conductivity $\kappa_P$, which follows the $T^{-1}$ decay typical of Peierls’ particle-like transport in crystals with dominant third-order anharmonicity. Blue, coherences conductivity $\kappa_{C}$, significant at high temperature. Black, total conductivity, $\kappa_{T}{=}\kappa_{P}{+}\kappa_{C}$. Scatter points are experiments from Aibing et al.  \cite{https://doi.org/10.1111/j.1551-2916.2009.03244.x}, Hongying et al.  \cite{Dong_2019}, Chenglong et al.  \cite{ZHANG20188818}, and Shijina et al.  \cite{shijina_very_2016}.
The theoretical conductivities are the average trace of the respective tensors, these are estimators for the conductivity of the polycrystalline samples employed in experiments  \cite{Schulgasser_1977}.}\label{TC-LaPO4} 
\end{figure}

In this section, we use the first-principles microscopic vibrational properties of LaPO$_{4}$ to evaluate the thermal conductivity as a function of temperature (Eq.~(\ref{eq:thermal_conductivity_combined})).
Fig.~\ref{TC-LaPO4} shows our theoretical predictions for a bulk sample (\textit{i.e.}, without considering grain-boundary scattering, more on this later); these are compared with experiments by Hongying et al.  \cite{Dong_2019}, Aibing et al.  \cite{https://doi.org/10.1111/j.1551-2916.2009.03244.x}, Chenglong et al.  \cite{ZHANG20188818}, and Shijina et al.  \cite{shijina_very_2016}.
Theory and experiments are in overall good agreement (more details on the spread of the experimental data will be discussed later), both approaching a $T^{-1}$ trend at low temperature and a milder-than-$T^{-1}$ decay at high temperature.
These different trends in the low and high temperature limits emerge from the coexistence of particle-like and wave-like transport mechanism, whose relative strength depends on temperature. Specifically, Fig.~\ref{TC-LaPO4} shows that the particle-like mechanisms---which determine the populations conductivity $\kappa_P$ discussed in Sec.~\ref{sec:Wigner_formulation}---dominate at low temperature.  This is a consequence of the increase of $\kappa_P(T)$ upon lowering temperature $T$, which can be understood recalling that 
the particle-like conductivity emerging from the Wigner formulation is totally equivalent to the LBTE conductivity \cite{Wigner_paper},  
and in crystals characterized by dominant third-order anharmonicity the latter universally follows a $\kappa_P(T)\sim T^{-1}$ scaling for $T>T_D$, where $T_D$ is the Debye temperature\cite{ziman_electrons_2001,sun_lattice_2010,PhysRevB.91.144304,lory_direct_2017}. For LaPO$_4$, $T_D\sim 144$ K\footnote{Fig.~\ref{phonon-disp} shows that in LaPO$_4$ the phonon modes approximatively have linear dispersion relation up to about $\hbar\omega_D\simeq$100 cm$^{-1}$. These modes give the largest contribution to $\kappa_P$, and their Debye temperature is approximatively $T_D= \frac{\hbar\omega_D}{k_B}\approx 144 $ K.}, explaining why $\kappa_P(T)\propto T^{-1}$ for T>200 K. 
In contrast, 
wave-like mechanisms yield a coherences conductivity $\kappa_C$ that in LaPO$_{4}$ increases with temperature and determines the milder-than-$T^{-1}$ decay at high temperature. 

Now we focus on the spread displayed by different, independent experiments \cite{Dong_2019,https://doi.org/10.1111/j.1551-2916.2009.03244.x,ZHANG20188818,shijina_very_2016}, which is deemed to originate from differences in the samples. Specifically, the conductivity is sensitive to the sample's grains properties and size  \cite{https://doi.org/10.1111/j.1551-2916.2010.03779.x,Zhang201627,Dong_2019}, which are determined by the synthesis process.
The experiments that are closest to our bulk (perfect-crystal) calculations are those by Hongying et al. \cite{Dong_2019}, who used heat treatment to obtain samples with a high degree of crystallinity (average grain size $\gtrsim 25$ $\mu m$). 
We also note that the highest-temperature experiment performed by Hongying et al.  \cite{Dong_2019} ($T\sim 1500 K$) departs from the decreasing trend displayed by all the other experiments discussed by the same reference. This change in trend might originate from the onset of the radiative conduction, the description of this effect goes beyond the scope of the present study.
The other experiments reported in Fig.~\ref{TC-LaPO4} show a lower conductivity, which originates from the smaller average grain size and consequent stronger grain-boundary scattering in these samples.
\begin{figure}[t]
\includegraphics[width=\WidthFigure]{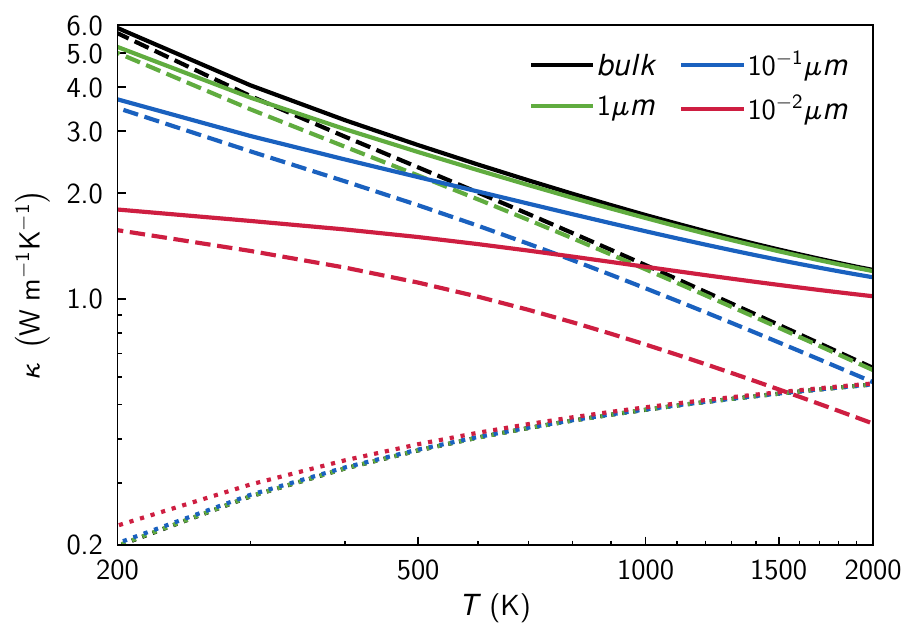}\caption{
\textbf{Effect of grain-boundary scattering on the conductivity.} We considered grain sizes equal to $10^{-2}$ (red),  $10^{-1}$ (blue) and $ 1  \thinspace \mu m$ (green) to compute the conductivity, and we compared results with the bulk value (black). The solid, dashed and dotted lines are $\kappa_{T}, \thinspace \kappa_{P}$ and $\kappa_{C}$, respectively.
 Each line is the average trace of the respective conductivity tensor, an estimator for the conductivity of the polycrystalline samples employed in experiments  \cite{Schulgasser_1977}.}\label{TC-prist-GB}
\end{figure}
Specifically, the samples used by Shijina et al.  \cite{shijina_very_2016} had grains with size of 1-4 $\mu m$, Aibing et al.  \cite{https://doi.org/10.1111/j.1551-2916.2009.03244.x} {used samples with grains in the {1-3} $\mu m$ range}, and Chenglong et al.  \cite{ZHANG20188818} {used samples with grains in the 2-5 $\mu m$ range}. We show in Fig.~\ref{TC-prist-GB} that accounting for grain-boundary scattering at the micrometer length scale (see Eq.~(\ref{eq:lw_bnd}) in Appendix~\ref{sec:contributions_to_phonon_linewidths}) produces variations of the total thermal conductivity that are broadly compatible with the spread observed in different experiments. Finally, we note that Fig.~\ref{TC-prist-GB} also reports predictions for samples with nanometre-sized grains (blue and red curves); this is to provide information on how much the conductivity would change if the experimental nanostructuring techniques for TBC (see e.g. Refs.~\cite{li_preparation_2017,guo_preparation_2017}) were used to prepare the samples.

\subsection{Particle-like \& wave-like transport}
\begin{figure*}
\begin{centering}
\includegraphics[scale=0.5]{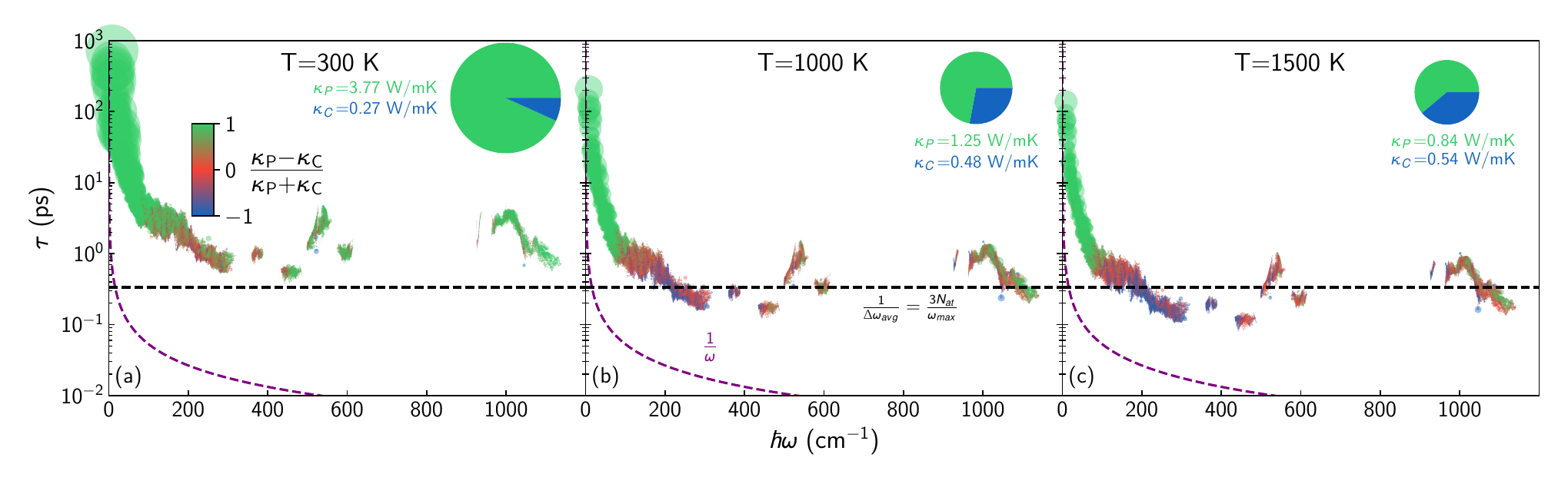}
\par\end{centering}
\begin{centering}
\includegraphics[scale=0.5]{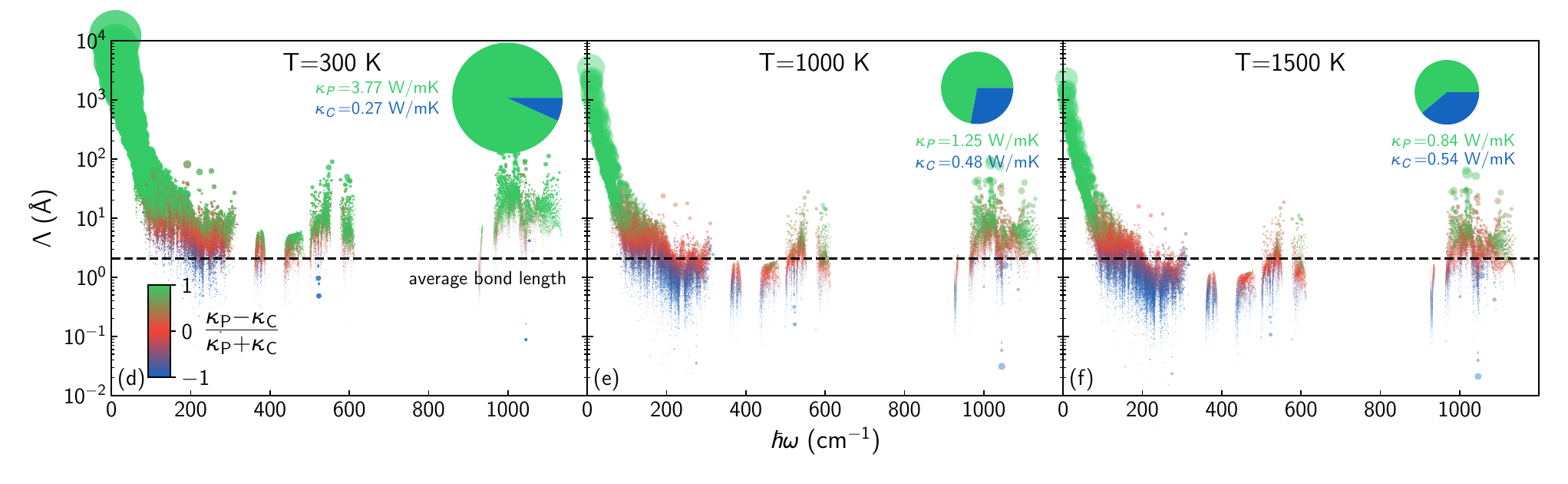}
\par\end{centering}
\caption{\textbf{Phonon lifetime (top) and mean free path (bottom) as a function of energy at different temperatures.} A color code has been used to show the origin of the conduction mechanism; green, blue, and red color represent the particle-like, wavelike, and mixed conductivity (50 $\%$ each), respectively. The area of each circle corresponds to its contribution to the total conductivity. In the upper panels, the horizontal black-dashed line is the "Wigner limit in time", where phonon lifetime, $\tau\bm{(q)}_{s}$ is equal to the inverse of average interbands spacing ($\tau\bm{(q)}_{s}=[\Delta\omega_{\rm avg}]^{-1}$, see text). The purple-dashed hyperbola $\tau=\frac{1}{\omega}$ indicates the regime of validity of the Wigner formulation, which requires phonons to be above such line to be well-defined (non-overdamped) quasiparticles  \cite{landau1980statistical}. 
 The horizontal dashed line in the mean free path panel is the average bond length of LaPO${_4}$. The pie charts have an area proportional to the total conductivity, the green and blue slices represent the particle-like and wave-like contributions, respectively. 
\label{Lifetime-MFP-plot}
 }
\end{figure*}

The calculations in the previous section highlighted how the macroscopic thermal conductivity is determined by both particle-like and wave-like microscopic transport mechanisms. 
In this section, we investigate these microscopic transport mechanisms; specifically, how their relative strengths and contributions to the total macroscopic conductivity vary as a function of temperature. As discussed in Ref.~\cite{Wigner_paper}, it is possible to resolve how much each phonon $(\bm{q})_s$ contributes to the particle-like ($\bar{\mathcal{K}}_{P}(\bm{q})_{s}$) and wave-like conductivities ($\bar{\mathcal{K}}_{C}(\bm{q})_{s}$) (the bar denotes the average trace of the mode-resolved contributions to the particle-like and wave-like conductivity tensors, full expressions are reported in Appendix~\ref{kp_and_kc}). Ref.~\cite{Wigner_paper} demonstrated that the relative strength of these contributions scales as the ratio between the average interband spacings  ($\Delta\omega_{\rm avg} {=} \frac{\omega_{\rm max}}{3N_{at}}$, where  $\omega_{\rm max}$ is the maximum phonon frequency, $N_{at}$ is the number of atoms in the primitive cell, and $3N_{at}$ the number of phonon bands) and the linewidth
$\Gamma(\bm{q})_{s}{=}[\tau(\bm{q})_{s}]^{-1}$ (here $\tau(\bm{q})_{s}$ is the phonon lifetime). In formulas,
\begin{equation}\label{eq1}
  \frac{\bar{\mathcal{K}}_{C}(\bm{q})_{s}}{\bar{\mathcal{K}}_{P}(\bm{q})_{s}} \simeq \frac{\Gamma(\bm{q})_{s}}{\Delta\omega_{\rm avg}} = \frac{[\Delta\omega_{\rm avg}]^{-1}}{\tau(\bm{q})_{s}}.
\end{equation}
Eq.~(\ref{eq1}) predicts that phonons having a lifetime $\tau(\bm{q})_s$ equal to the inverse interband spacing $[\Delta\omega_{\rm avg}]^{-1}$ (also referred to as ``Wigner limit in time''  \cite{Wigner_paper}) contribute simultaneously and with equal strength to both particle-like and wave-like conduction mechanism.
In contrast, phonons with a lifetime much longer (shorter) than the Wigner limit in time contribute predominantly to the particle-like (wave-like) conductivity. 
Finally, Eq.~(\ref{eq1}) predicts the transition between these two limits to be non-sharp and centered at the Wigner limit in time. 
These analytical expectations are verified numerically in Fig.~\ref{Lifetime-MFP-plot}. Specifically, in the upper panel of such figure, we show the distribution of phonon lifetimes as a function of phonon energies at different temperatures.  
For each phonon mode (individual scatter point in Fig.~\ref{Lifetime-MFP-plot}), we use the particle-like and wave-like conductivity contributions ($\bar{\mathcal{K}}_{P}(\bm{q})_{s}$ and $\bar{\mathcal{K}}_{C}(\bm{q})_{s}$ appearing in Eq.~(\ref{eq1}), respectively) to resolve the conduction mechanisms through which the phonon participates to heat transport, as well as how much the microscopic phonon mode contributes to the macroscopic conductivity.
The first information on the type of conduction mechanisms is encoded in the color of the scatter point, determined according to the value of the parameter
 \begin{equation}\label{eq3}
   c=\frac{\bar{\mathcal{K}}_{P}(\bm{q})_{s}-\bar{\mathcal{K}}_{C}(\bm{q})_{s}}{\bar{\mathcal{K}}_{P}(\bm{q})_{s}+\bar{\mathcal{K}}_{C}(\bm{q})_{s}}.
\end{equation} 
Eq.~(\ref{eq3}) implies that $c{=}{+}1$ when the phonon $(\bm{q})_s$ predominantly contributes to particle-like conduction (corresponding to green),  $c{=}{-}1$ when instead the phonon contributes mainly to wave-like conduction (blue), and finally $c{=}0$ when the phonon contributes equally to particle-like and wave-like conduction (red).
In the following, we employ a linear color scale interpolating blue, red, and green to resolve possible intermediate cases. The second information, on the magnitude of the transport mechanisms, is represented by the area of each scatter point, which is proportional to the contribution of such phonon to the total thermal conductivity (size $\propto\bar{\mathcal{K}}_{P}(\bm{q})_{s}{+}\bar{\mathcal{K}}_{C}(\bm{q})_{s}$). 
Fig.~\ref{Lifetime-MFP-plot} shows that the relative strength of particle-like and wave-like mechanisms strongly depends on the energy of the phonon, and varies significantly with temperature. Specifically, at room temperature (panel a), most of the phonons have a lifetime $\tau(\bm{q})_s$ above the Wigner limit in time $[\Delta\omega_{\rm avg}]^{-1}$
(represented by the horizontal dashed-black line), and their green color indicates that they mainly contribute to particle-like transport. Increasing temperature (panel b is 1000 K, and panel c is 1500 K) yields a reduction of the lifetimes;  numerical results confirm the analytical expectations that phonons with a lifetime comparable to the Wigner limit in time (red) contribute simultaneously to particle-like and wave-like conduction mechanisms, and phonons with even shorter lifetime mainly contribute to wavelike transport. These findings can be intuitively understood recalling that phonons with an extremely short lifetime are suppressed very quickly and thus cannot propagate particle-like enough to yield a sizable contribution to the populations conductivity. However, these short-lived phonons can still interfere and tunnel wavelike---we recall that interference and tunneling occur also between damped waves---resulting in a significant contribution to the wave-like conductivity.  Finally, we highlight that all the phonons in LaPO$_{4}$ have a lifetime longer than their reciprocal frequency ($\tau(\bm{q})_s>[\omega(\bm{q})_s]^{-1}$, \textit{i.e.} they are all above the dashed-purple line in panels a,b,c); thus Landau's quasiparticle picture  \cite{landau1980statistical} for phonons holds in LaPO$_{4}$ and consequently the Wigner formulation can be applied  \cite{Wigner_paper,caldarelli_many-body_2022}.

The lifetime-energy analysis reported in the upper panel of Fig.~\ref{Lifetime-MFP-plot} sheds light on the microscopic timescales underlying heat transport, and how they affect the macroscopic conductivity. In order to gain further insights, particularly on the dependence of the conductivity on the grains' length scales discussed in the previous section, it is useful to recast such an analysis in the space-energy domain. To achieve this, we multiply the phonon lifetimes by the corresponding group velocities, obtaining the microscopic propagation length scales of phonons (mean free paths, MFP) $\Lambda(\bm{q})_s{=}\frac{1}{\sqrt{3}}|\!|\tens{v}(\bm{q})_{ss}|\!|\tau(\bm{q})_s$ (here, $\frac{1}{\sqrt{3}}|\!|\tens{v}(\bm{q})_{ss}|\!|$ is the spatially averaged modulus of the group velocity).
 
The bottom panels of Fig.~\ref{Lifetime-MFP-plot} show the MFP vs phonon energy at 300 K (a), 1000 K (b), and 1500 K (c). Similarly to the phonon lifetime-energy plots, a crossover from particle-like to wave-like transport is clearly evident here as well, and such a non-sharp transition is centered around the average bond length  (see Ref.~\cite{Wigner_paper} for details on the relation between the particle-wave crossover in space and the average bond length). We highlight how phonons with MFP larger (smaller) than the average bond length contribute to particle (wave-like) conduction, and phonons with MFP equal to the average bond length contribute simultaneously to both particle and wave mechanisms. Finally, we note that from the lower panels of Fig.~\ref{Lifetime-MFP-plot} it is apparent that most of the phonons in LaPO$_4$ have MFP always equal or shorter than one micrometer, rationalizing the small difference between the bulk thermal conductivity and the thermal conductivity computed accounted for grain-boundary scattering at length scale $1\mu m$ discussed in Fig.~\ref{TC-prist-GB}.

\subsection{Engineering the thermal conductivity through compositional disorder}\label{comp-disorder}
\subsubsection{Effects of compositional disorder on vibrational properties} 
\label{sub:mass_sub}
 
La$_{1-x}$Gd$_{x}$PO$_{4}$ alloys are promising materials for future thermal barrier applications, due to their excellent thermal stability and chemical durability  \cite{NEUMEIER2017396}. In general, the presence of compositional disorder (alloying) causes a reduction of the thermal conductivity  \cite{garg_thermal_2011,thebaud_success_2020,thebaud_perturbation_2022,braun_charge-induced_2018,padture_advanced_2016,clarke_thermal-barrier_2012,cahill_nanoscale_2014,seyf_rethinking_2017, R-Farris-PRB98_220301_2018, Pandey_2017, Zlatan_PRA_2016}, thus is expected to be beneficial for TBC applications. 
To the best of our knowledge, there are no theoretical or experimental works on the thermal conductivity of La$_{1-x}$Gd$_{x}$PO$_{4}$ alloys. 
As anticipated in the introduction, recent work  \cite{thebaud_perturbation_2022} has highlighted how the standard LBTE-based perturbative treatment of compositional disorder  \cite{garg_thermal_2011,PhysRevB.27.858,PhysRevLett.106.045901} --- which is accurate in weakly anharmonic systems in which low-frequency vibrational modes dominate transport  \cite{thebaud_success_2020,PhysRevLett.106.045901,garg_thermal_2011} --- fails in systems where transport is not dominated by low-frequency modes. We have seen in Fig.~\ref{Lifetime-MFP-plot} that LaPO$_4$ at high temperature belongs to this class, hence in this section, we develop a computational protocol that exploits the Wigner formulation to describe compositional-mass disorder explicitly, overcoming the limitations of the standard LBTE-based perturbative treatment of disorder. 

To describe how compositional-mass disorder affects the conductivity in La$_{1-x}$Gd$_{x}$PO$_{4}$ alloys, we need to compute the quantities entering in Eq.~(\ref{eq:thermal_conductivity_combined})---\textit{i.e.}, vibrational frequencies, velocity operators, and linewidths---in compositionally disordered structures.
Starting from the harmonic frequencies and velocity operators, we compute them explicitly in the mass-substitution approximation \cite{RevModPhys.73.515} according to the following procedure: (i) we build a perfect supercell (of size up to $6{\times}6{\times}6$, \textit{i.e.}, containing up to 5184 atoms) of the primitive cell of LaPO$_{4}$; (ii) we destroy the crystalline order within such supercell by introducing compositional disorder (hereafter we will use `disordered cell' to refer to this compositionally disordered atomic structure); (iii) we calculate the vibrational properties of the disordered cell.
In particular, in step (ii) we introduce compositional disorder in the rare-earth site by replacing the mass of La with that of Gd with probability $x\in[0,1]$ in the mass-rescaled force-constant tensor:
\begin{equation}
	\tenscomp{G}_{\bm{R}b\alpha,\bm{R'}b'\hspace*{-0.5mm}\alpha'}=
	\frac{1}{\sqrt{f_x(M_b)f_x(M_{b'})}}\frac{\partial^2 {V} }{\partial {u}(\bm{R})_{b\alpha} \partial {u}(\bm{R}')_{b'\hspace*{-0.5mm}\alpha'} }\Big\lvert_{\rm eq},
	\label{eq:matrix_G}
\end{equation}
where $\bm{R}$ is a Bravais vector, $b$ denotes an atom having mass $M_b$ and position in the primitive cell $\bm{\tau}_b$, and $\alpha$ is a Cartesian direction; $\frac{\partial^2 {V} }{\partial {u}(\bm{R})_{b\alpha} \partial {u}(\bm{R}')_{b'\hspace*{-0.5mm}\alpha'} }\Big\lvert_{\rm eq}$ is the second derivative of the Born-Oppenheimer potential evaluated at equilibrium atomic positions, which within the mass-substitution approximation\cite{RevModPhys.73.515} is considered to be equal to that of pure LaPO$_{4}$.
The function $f_x$ introduces compositional-mass disorder on the La sites by  
replacing the mass of La with that of Gd with probability $x\in[0,1]$:
\begin{equation}
f_x(M_{\rm La})=\bigg\{\begin{array}{l}
	M_{\rm La} \text{  with probability  } 1-x;\\
	M_{\rm Gd} \text{ with probability } x;
\end{array}	
\label{eq:fun_repl}
\end{equation}
while it leaves unaffected the masses of O and P ($f_x(M_{\rm O})=M_{\rm O}$ and $f_x(M_{\rm P})=M_{\rm P}$ $\forall \;x$). 
Then, the disordered harmonic mass-rescaled force-constant tensor~(\ref{eq:matrix_G}) is used to compute the  dynamical matrix at wavevector $\bm{q}$,
\begin{equation}
  \tenscomp{D}(\bm{q})_{b\alpha,b'\!\alpha'}{=}\sum_{\bm{R}}\tenscomp{G}_{\bm{R}b\alpha,\bm{0}b'\!\alpha'}e^{-i\bm{q}\cdot(\bm{R}{+}\bm{\tau}_b{-}\bm{\tau}_{b'})},
  \label{eq:dynamical}
\end{equation}
which is then diagonalized
\begin{equation}
  \sum_{b'\alpha'}\tenscomp{D}(\bm{q})_{b\alpha,b'\!\alpha'}\mathcal{E}(\bm{q})_{s,b'\!\alpha'}=\omega^2(\bm{q})_s\mathcal{E}(\bm{q})_{s,b\alpha}.
  \label{eq:diagonalization_dynamical_matrix}
\end{equation}
The eigenvalues appearing in Eq.~(\ref{eq:diagonalization_dynamical_matrix}) are related to the vibrational frequencies $\omega(\bm{q})_s$, and the eigenvectors  $\mathcal{E}(\bm{q})_{s,b\alpha}$ describe how atom $b$ moves along the Cartesian direction $\alpha$ when the phonon with wavevector $\bm{q}$ and mode $s$ is excited.
As discussed in Ref.~\cite{Wigner_paper}, the velocity operator is obtained from these quantities as
\begin{equation} 
\begin{split}     
 \hspace*{-3mm}\tenscomp{v}^\beta(\bm{q})_{s,s'}{=}\!\!\sum_{\substack{b,\alpha,b'\!,\alpha'}}\!\!\!\mathcal{E}^\star(\bm{q})_{{s},b\alpha}{{\nabla^{\beta}_{\bm{q}} \sqrt{\tenscomp{D}(\bm{q})}} }_{b\alpha,b'\hspace*{-0.5mm}\alpha'} \mathcal{E}(\bm{q})_{s'\hspace*{-0.5mm},b'\hspace*{-0.5mm}\alpha'}.
\end{split}    
   \raisetag{10mm} 
\label{eq:vel_op}
\end{equation}

To estimate the effect of compositional disorder on the linewidths, we focus on bulk systems with negligible (zero) grain-boundary linewidths. In addition, we neglect intrinsic isotopic disorder, since it has negligible effects on the conductivity of LaPO$_{4}$ (see Fig.~\ref{fig:negligible_ISO} in Appendix).
Thus, we use the mass-substitution approximation to estimate the effect of compositional disorder on the dominant, anharmonic part of the linewidths, $\hbar\Gamma^{\rm anh}(\bm{q})_s$. Specifically, we evaluate $\hbar\Gamma^{\rm anh}(\bm{q})_s$ (Eq.~(\ref{eq:anh_linewidth}) in Appendix) using third-order force constants modified to account for compositional disorder using the aforementioned mass-replacement function $f(M_b)$ (see Eq.~(\ref{eq:third_order_matrix_element}) in Appendix).

We show in Fig.~\ref{TC-LaPO4-GDPO4} that the mass-substitution approximation captures the lower conductivity of pure GdPO$_{4}$ compared to pure LaPO$_{4}$, in agreement with experiments (see Refs.~\cite{https://doi.org/10.1111/j.1551-2916.2009.03244.x,winter_oxide_2007} for the former, and Refs.~\cite{https://doi.org/10.1111/j.1551-2916.2009.03244.x,Dong_2019,ZHANG20188818,shijina_very_2016} for the latter). 
Therefore, the mass-substitution approximation is accurate enough for the scope of the present analysis.
The inset of Fig.~\ref{TC-LaPO4-GDPO4} shows that GdPO$_{4}$ has conductivity lower than LaPO$_{4}$ because of a simultaneous reduction of both the particle-like and wave-like conductivities. The reduction of $\kappa_P$ caused by the substitution La$\to$Gd can be understood from the increase in the linewidth shown in Fig.~\ref{LW_fit-TC-compare}, and the reduction of the energy of the acoustic modes (\textit{i.e.}, reduction in the phonon group velocities) shown in Fig.~\ref{Alloy-config-DOS}\textbf{c}. The reduction of $\kappa_C$ upon La$\to$Gd substitution is caused by the decrease of the off-diagonal velocity-operator elements, which accompanies the decrease of the group velocities (diagonal velocity-operator elements).

\begin{figure}
\includegraphics[width=\WidthFigure]{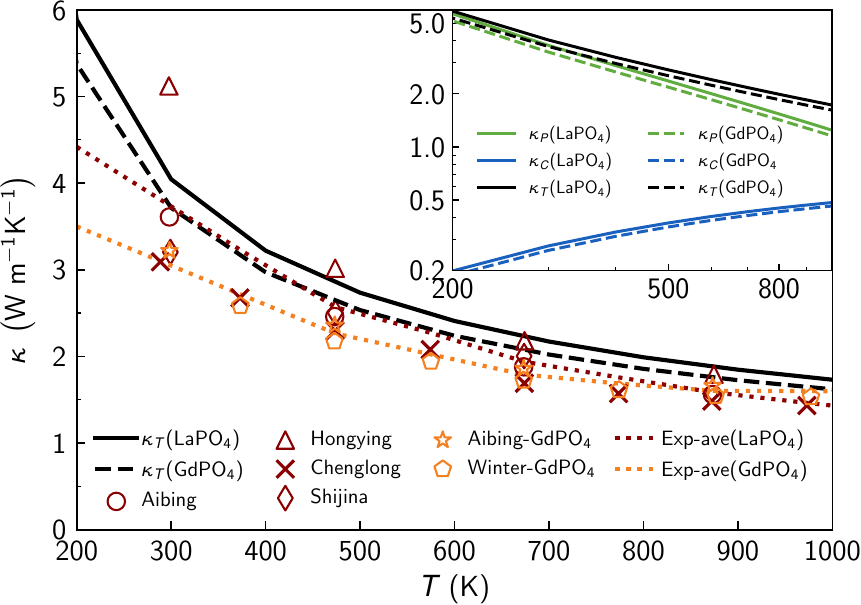}
\caption{\textbf{Conductivities of pure LaPO$_4$ and GdPO$_4$.} 
The total conductivity of LaPO$_4$ (GdPO$_4$) is solid (dashed) black, and experiments are red\cite{https://doi.org/10.1111/j.1551-2916.2009.03244.x,Dong_2019,ZHANG20188818,shijina_very_2016} (orange\cite{https://doi.org/10.1111/j.1551-2916.2009.03244.x,winter_oxide_2007}) symbols. The dotted red (orange) line is the average of the experimental data for LaPO$_4$ (GdPO$_4$). Inset, the reduction of total conductivity  ($\kappa_T$)  caused by completely replacing La$\to$Gd originates from a simultaneous reduction of particle-like ($\kappa_P$) and wave-like ($\kappa_C$) conductivities.
}\label{TC-LaPO4-GDPO4} 
\end{figure}

The knowledge of the anharmonic linewidths of pristine LaPO$_{4}$ and GdPO$_{4}$ (see Fig. \ref{LW_fit-TC-compare} in Appendix, both these pristine materials have a 24-atom primitive cell), and the unaffordable computational cost for calculating of the anharmonic linewidths in disordered cells of La$_{1-x}$Gd$_x$PO$_{4}$  containing thousands of atoms\cite{simoncelli_thermal_2023}, motivate us to adopt the approximate description of anharmonic linewidths for alloys introduced in past work\cite{simoncelli_thermal_2023,thebaud_perturbation_2022,garg_thermal_2011}. In practice, this corresponds to: (i) coarse-graining the frequency-linewidths distributions of pure LaPO$_{4}$ and pure GdPO$_{4}$ into single-valued functions of $\omega$, $\Gamma^{\rm a}_{\rm LaPO_4}(\omega)$ and $\Gamma^{\rm a}_{\rm GdPO_4}(\omega)$ (Appendix~\ref{sec:accounting_for_anharmonicity_at_a_reduced_computational_cost} discusses the Eq.~(\ref{eq:approx_analytical_f}) used for such coarse-graining, and the accuracy of the approximations performed); (ii) determining the linewidths of the disordered alloy by  composition-weighted average of the LaPO$_{4}$ and GdPO$_{4}$ linewidths\cite{Linear_intpol_Katcho}:
\begin{equation}
	\Gamma^{\rm a}_{\rm La_{1-x}Gd_xPO_4}(\omega)=(1-x)\Gamma^{\rm a}_{\rm LaPO_4}(\omega)+x\;\Gamma^{\rm a}_{\rm GdPO_4}(\omega).
	\label{eq:mix_lw}
\end{equation}
\subsubsection{Effects of compositional disorder on thermal properties}  
\label{ssub:comp_dis_vib_prop}

After having computed the harmonic frequencies and velocity operators, and anharmonic linewidths in a compositionally disordered cell with Eq.~(\ref{eq:mix_lw}), we use these quantities to evaluate the thermal conductivity~(\ref{eq:thermal_conductivity_combined}). Implementation details of this protocol are provided in Appendix~\ref{sub:thermal_conductivity}.

\begin{figure}[b]
\begin{centering}
\includegraphics[width=\WidthFigure]{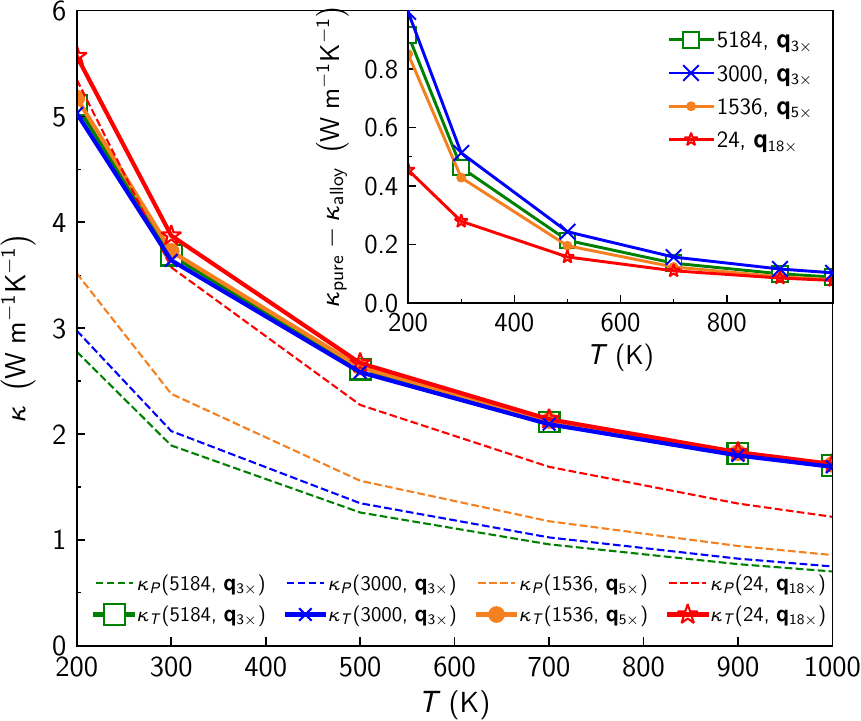}
\par\end{centering}
\caption{\textbf{Thermal conductivity of models of La$_{0.5}$Gd$_{0.5}$PO$_{4}$ having different sizes.} The solid lines with the markers are total conductivities $\kappa_{T}$, dashed lines are the particle-like conductivities $\kappa_{P}$.
Different colors denote calculations done in disordered cells having different size and using different $\bm{q}$-meshes (in the legend, numbers represent the number of atoms in the disordered cells, and `$\bm{q}_{n\times}$' means that conductivities are evaluated on a $n{\times}n{\times}n$ $\bm{q}$ mesh). $\kappa_{P}$ decreases as the size of the model increases; accounting also for the wave-like conductivity $\kappa_{C}$ allows, in models containing 1536 or more atoms, to converge to the bulk (size-independent) limit of the total conductivity $\kappa_{T}=\kappa_{P}+\kappa_{C}$.
Inset, difference between the total conductivity of pure LaPO$_4$, and that of La$_{0.5}$Gd$_{0.5}$PO$_{4}$ described with disordered cells containing 24, 1536, 3000, and 5184 atoms. 
}\label{supecercell-convergence}
\end{figure}

\begin{figure}[b]
	\centering
	\includegraphics[width=\columnwidth]{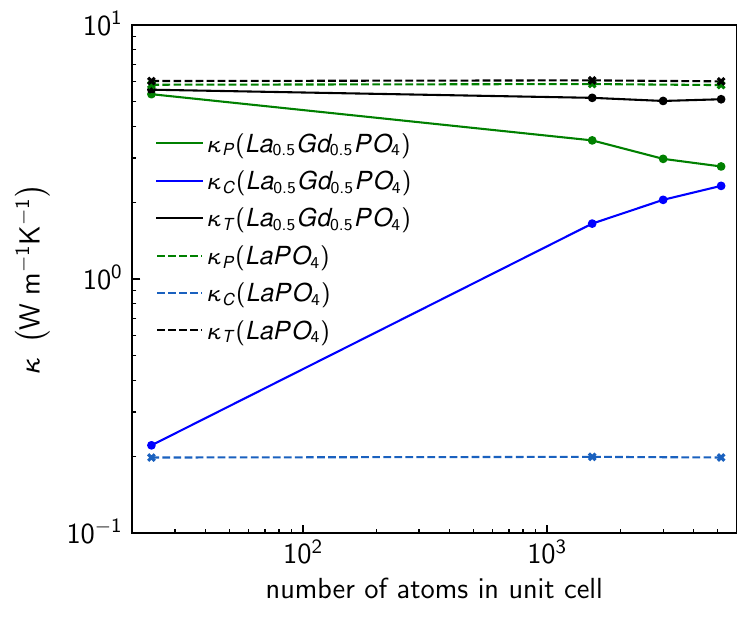}
	\caption{\textbf{Necessity to consider both $\kappa_P$ and $\kappa_C$ to describe the thermal conductivity of alloys.}
	Green, blue, and black are particle-like ($\kappa_P$), wave-like ($\kappa_C$), and total ($\kappa_T{=}\kappa_P{+}\kappa_C$) conductivities at 200 K for pristine LaPO$_4$ (dashed), and La$_{0.5}$Gd$_{0.5}$PO$_4$ alloy (solid). We highlight how when the alloy is accurately described with large disordered cells, the total conductivity $\kappa_T$ respects the physical expectation of being independent from the number of atoms in the alloy's unit cell, even if the ratio $\kappa_P/\kappa_C$ is size-dependent \cite{simoncelli_thermal_2023}.}
	\label{fig:alloy_k}
\end{figure}

The description of how compositional disorder affects thermal conductivity becomes more accurate as one increases the size of the disordered cell. In practice, one has to verify numerically that the disordered cell is large enough to yield computational convergence; this can be done by computing all the parameters entering into Eq.~(\ref{eq:thermal_conductivity_combined}) in increasingly larger disordered cells, and verifying that for sizes larger than a certain value the conductivity remains practically unchanged. 
In Fig.~\ref{supecercell-convergence} we test the computational convergence for the case $x=0.5$, where we show that employing disordered cells of size 1536, 3000, and 5184 atoms yields practically indistinguishable results for the total conductivity. 
It is worth noting that in these calculations we used Fourier interpolation to improve the accuracy with which bulk vibrational and thermal properties are extrapolated from the finite atomistic model\cite{allen_thermal_1993,feldman_thermal_1993,simoncelli_thermal_2023}. 
In particular we computed these properties using a $5{\times}5{\times}5$ $\bm{q}$ mesh for the 1536-atom cell, a $3{\times}3{\times}3$ $\bm{q}$ mesh for the 3000-atom cell, and a $3{\times}3{\times}3$ $\bm{q}$ mesh for the 5184-atom cell. 
We recall that performing a Fourier interpolation on a $n{\times}n{\times}n$ $\bm{q}$ mesh spanning the Brillouin zone of a certain reference cell allows to simulate a system that is $n{\times}n{\times}n$ larger, but with a disorder length scale limited to the size of the reference cell \cite{simoncelli_thermal_2023,harper_vibrational_2023}\footnote{For example, for a 1536-atom model, a calculation on a $5{\times}5{\times}5$ $\bm{q}$-mesh corresponds to simulating a system that contains $1536\cdot 5^3=192000$ atoms, but with disorder length scale limited to the size of the reference cell containing 1536 atoms.}.
Therefore, comparing calculations performed using disordered cells of different sizes, and using different meshes, provides information about how much the conductivity is affected by finite-size effects.
Fig.~\ref{supecercell-convergence} shows that the total conductivities obtained from disordered atomistic models having cell size (\textit{i.e.}, disorder length scales) ranging from 1536 to 5184 atoms are practically indistinguishable in the $\bm{q}$-sampling limit\footnote{With `$\bm{q}$-sampling limit' we mean the regime in which increasing the $\bm{q}$ mesh yields negligible changes on the conductivity}; this implies that, for La$_{1-x}$Gd$_x$PO$_4$ alloys, disordered cells containing 1536 atoms are already sufficiently large to yield a computationally converged value for the thermal conductivity.

Fig.~\ref{supecercell-convergence} also highlights that, in order to predict the bulk limit of the thermal conductivity of an alloy, it is necessary to: (i) account for both particle-like and wave-like transport mechanisms via Eq.~(\ref{eq:thermal_conductivity_combined}); (ii) evaluate Eq.~(\ref{eq:thermal_conductivity_combined}) using compositionally disordered cells of size sufficiently large to accurately describe disorder. 
These two requirements are related. In fact, introducing compositional disorder in LaPO$_4$ causes a reduction of $\kappa_P$\footnote{The decrease of particle-like conductivity upon increasing disorder was already noted in Ref.~\cite{garg_thermal_2011} for Si$_{0.5}$Ge$_{0.5}$ alloys. This can be intuitively understood recalling that disorder induces repulsion between vibrational eigenstates, lowering the phonon group velocities\cite{simkin_minimum_2000}, and consequently lowering the particle-like thermal conductivity.} and an increase of $\kappa_C$. 
We find that in La$_{1-x}$Gd$_x$PO$_4$ alloys, upon increasing the size of the disordered cell (\textit{i.e.}, accuracy  in the description of disorder) beyond 1536 atoms, the total Wigner conductivity $\kappa_T{=}\kappa_P{+}\kappa_C$ converges to a size-independent, bulk limit of the alloy's conductivity (inset of Fig.~\ref{supecercell-convergence}, and Fig.~\ref{fig:alloy_k}).

As discussed already in Refs.~\cite{allen_thermal_1993,garg_thermal_2011}, in the presence of disorder the particle-like conductivity depends on the size of the disordered cell used to describe the system, and converges to zero in the limit of infinitely large disordered cell. This last statement is supported by numerical evidence in Fig.~\ref{fig:alloy_k}, Fig.~4.7 of Ref.~\cite{garg_thermal_2011}, and Fig.~12 of Ref.~\cite{simoncelli_thermal_2023}, and additional details are provided in Appendix~\ref{sec:size_consistency_of_the_lbte_in_pristine_systems}.
To further confirm and validate that the reduction of the conductivity is caused by compositional disorder, we verified that in a pristine LaPO$_4$ crystal without disorder $\kappa_P$, $\kappa_C$, and $\kappa_T$ are all independent from the arbitrary choice of the crystal's unit cell: the dashed lines in Fig.~\ref{fig:alloy_k} demonstrate this for cells containing 24 atoms (primitive cell), 1536 atoms (4$\times$4$\times$4 supercell), and 5184 atoms (6$\times$6$\times$6 supercell); Fig.~\ref{fig:LBTE_size_cons} in Appendix shows analogous tests performed at higher temperatures. 

After having validated the computational protocol to describe disorder in the mass-substitution approximation within the Wigner framework, we analyze how the variable compositional disorder affects thermal conductivity. 
\begin{figure}[t]
\includegraphics[width=\WidthFigure]{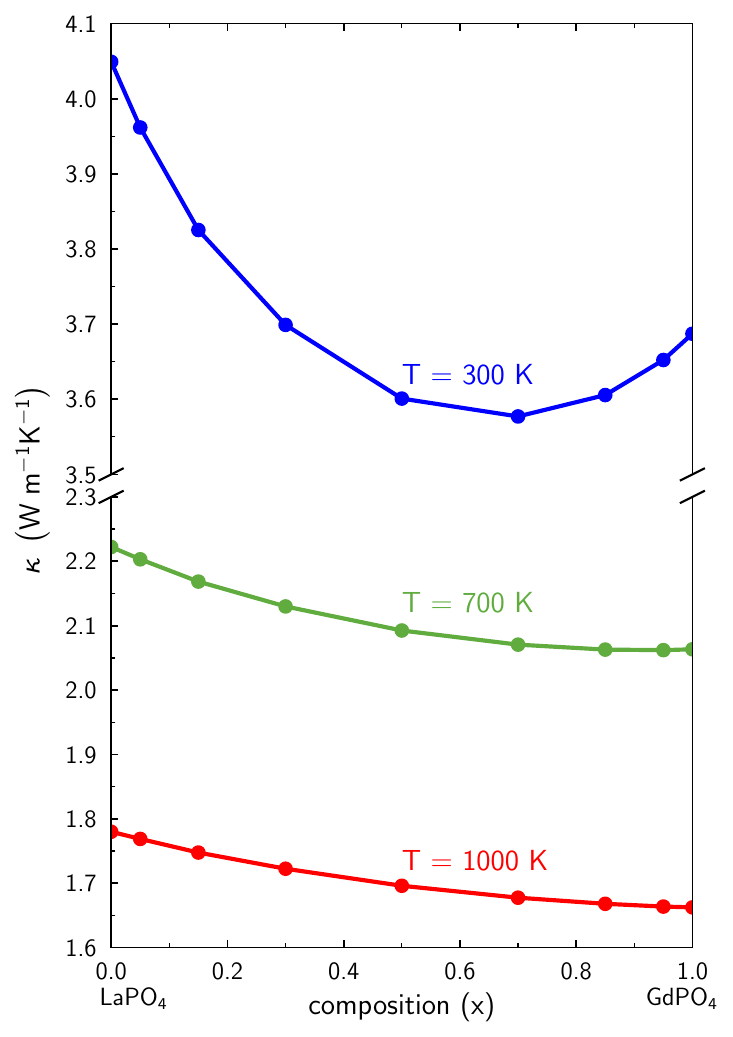}\caption{\textbf{Thermal conductivity of La$_{1-x}$Gd$_{x}$PO$_{4}$ alloys} as a function of composition, at 300, 700 and 1000 K. At 300 K, the conductivity drops as compositional disorder increases, reaching a minimum at $x\sim0.7$. The effect of compositional disorder becomes less relevant as temperature increases.}\label{TC-composition}
\end{figure}
Fig.~\ref{TC-composition} shows the conductivity of La$_{1-x}$Gd$_{x}$PO$_{4}$ alloys at different compositions and various temperatures. 
We see that, compared to pristine LaPO$_4$, the conductivity decreases as compositional disorder increases. Such a decrease is more pronounced at low temperatures (300 K), where anharmonicity is weak and does not dominate over disorder in limiting heat transfer. At 300K, the dependence of the conductivity from composition bears analogies with that observed in other alloys  \cite{PhysRevLett.106.045901, PhysRevB.86.174307, seyf_rethinking_2017, Colombo_PRL_112_065901, PhysRevB.98.115205, _rowe_thermoelectrics_handbook_2005, Abels_TC_disorder,wan_glass-like_2010, WAN_LaGdZrO-alloy_TC_PRB},
displaying a U-shaped trend which reaches a minimum at $x\sim 0.7$.
We note that in the minimum the conductivity is about 12\,\% lower than that of pristine LaPO$_4$.  Such a decrease is much smaller than the order-of-magnitude decrease observed in archetypal alloys such as Si$_{x}$Ge$_{1-x}$ \cite{PhysRevLett.106.045901}, because of two reasons. First, anharmonicity in LaPO$_4$ and GdPO$_4$ is much stronger than that in Si and Ge, yielding a much lower thermal conductivity already in the pristine components; second, there is less mass contrast between La and Gd (mass ratio, Gd/La = 1.13) compared to Si and Ge (mass ratio, Ge/Si =2.59). When increasing the temperature, anharmonic effects become stronger and thus progressively dominate over compositional disorder in determining the conductivity; this is apparent from the almost negligible dependence of the conductivity from compositional disorder (and disappearance of the U-shaped trend) observed at 1000 K. 

In summary, we have developed a computational protocol that allows us to evaluate the effect of compositional disorder within the Wigner framework. This protocol allows to shed light on how the interplay between compositional disorder and anharmonicity determines the conductivity, and will be potentially very useful to study materials for next-generation TBCs.

\subsubsection{Micrometer-scale disorder: LaPO$_{4}$/La$_2$Zr$_2$O$_7$ composites}
\label{subs:cont_Garnett}
\begin{figure}[t]
\includegraphics[width=\WidthFigure]{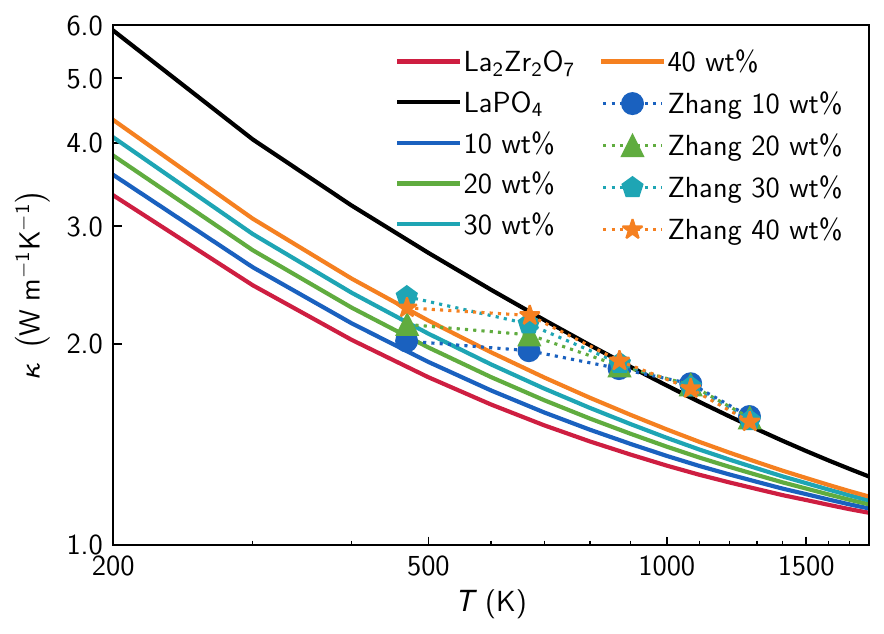}\caption{\textbf{Conductivity of LaPO${_4}$/La${_2}$Zr${_2}$O${_7}$ composites.} LaPO${_4}$ (filler) is added to La${_2}$Zr${_2}$O${_7}$ (matrix) at different weight percentages, and the thermal conductivity is computed using Maxwell's effective-conductivity model  for composites \cite{sevostianov_maxwells_2019,maxwell_treatise_1873}.  Experimental data are taken from Zhang et al.  \cite{Zhang201627}. \label{composite-TC} }
\end{figure}

La$_2$Zr$_2$O$_7$ has been identified as another potential TBC material with very low thermal conductivity \cite{liu_advances_2019}. However, its low thermal expansion coefficient causes the formation of cracks and delamination at high temperatures  \cite{CHEN2009391}. Composite structures of La$_2$Zr$_2$O$_7$ and LaPO$_{4}$ have been suggested to offer better thermomechanical properties compared to the base materials  \cite{Zhang201627}. Computing the thermal conductivity of composites characterized by compositional disorder at the micrometer length scale has a prohibitively high computational cost for first-principles methods. Therefore, in order to have a qualitative understanding of heat conduction in LaPO$_{4}$/La$_2$Zr$_2$O$_7$ composites, we employ the continuum Maxwell \cite{maxwell_treatise_1873} model for the effective  conductivity of a composite
(see Eq. 2.14 of Ref.~\cite{sevostianov_maxwells_2019}),    
which determines the total thermal conductivity of the composite ($\kappa$) from the total conductivities of the of matrix ($\kappa_1$) and filler ($\kappa_2$):
\begin{equation}
    \frac{\kappa-\kappa_1}{\kappa+2\kappa_1}={V}\frac{\kappa_2-\kappa_1}{\kappa_2+2\kappa_1},
    \label{eq:MG}
\end{equation}
where V is the volume fraction of the filler. Here we consider La$_2$Zr$_2$O$_7$ as the matrix (its conductivity is taken from ref  \cite{Wigner_paper}), and LaPO$_{4}$ as filler.
As in the experimental work by Zhang et al.,  \cite{Zhang201627}, we consider composites containing weight percentages LaPO$_{4}$ equal to 10, 20, 30, and 40 ${\%}$, and we estimate the conductivity of the composite using Eq.~(\ref{eq:MG}). Fig. \ref {composite-TC} shows the thermal conductivity of the composite as a function of temperature at different filler fractions. The conductivity increases with an increase in the fraction of LaPO$_4$; in all the cases, it is higher than that of pristine La$_2$Zr$_2$O$_7$ and smaller than that of LaPO$_4$. This trend is in broad agreement with the experiments performed by Zhang et al.  \cite{Zhang201627}. 
We also note that experimental samples have porosities that vary with the filler fraction, as well as interfaces between La$_2$Zr$_2$O$_7$ and LaPO$_{4}$ phases.
These effects are not accounted for by Maxwell's effective-conductivity model employed here, and their description might improve the agreement between theory and experiments in Fig.~\ref{composite-TC}, but it is an open challenging problem that is beyond the scope of the present paper.  

\section{Conclusions}\label{conclusions}
We have shown that the Wigner formulation of thermal transport \cite{Wigner_paper} can be combined with first-principles techniques to quantitatively predict the macroscopic thermal-insulation performance of compositionally disordered materials for thermal barrier coatings \cite{clarke_thermal-barrier_2012,zhang_al2o3-modified_2020,tejero-martin_review_2021,leng_solution_2022,NEUMEIER2017396}. First, we have investigated heat transport in pristine LaPO$_{4}$, showing that the recently developed
 Wigner transport equation \cite{simoncelli_unified_2019,Wigner_paper} rationalizes the conductivity decay milder than $T^{-1}$ observed in experiments \cite{https://doi.org/10.1111/j.1551-2916.2009.03244.x,Dong_2019,ZHANG20188818,shijina_very_2016}. More precisely, we have shown that the macroscopic trend of the conductivity is determined by the coexistence of microscopic particle-like and wave-like conduction mechanisms: the former strongly decays with temperature, while the latter increases with temperature, and the sum of the two yields the mildly decreasing trend observed in experiments.
We have discussed how grain-boundary scattering affects the conductivity of LaPO$_{4}$, showing that such a mechanism has a weak effect in samples having micrometric grains, but would become very important in nanostructured samples \cite{li_preparation_2017,guo_preparation_2017}. In particular, we have analyzed how the relative strength of the particle and wave-like transport mechanisms depends on the temperature, energy, and mean free path of the microscopic heat carriers.
We have developed and tested a computational protocol that allows to describe explicitly within the Wigner formulation how compositional disorder affects the conductivity, and we employed it to investigate the thermal properties of La$_{1-x}$Gd$_{x}$PO$_{4}$ alloys.
In particular, we described compositionally disordered samples  
containing up to 5184 atoms, a size that is one order of magnitude larger than that tractable by state-of-the-art first-principles molecular dynamics techniques. 
We discussed how the interplay between anharmonicity and disorder affects thermal transport in La$_{1-x}$Gd$_{x}$PO$_{4}$ alloys at different temperatures, showing that disorder has strong effects around room temperature and almost negligible effects at high temperatures ($\gtrsim$ 700 K).
We have gained two important insights on the problem of predicting the thermal conductivity of alloys with non-perturbative compositional disorder (\textit{i.e.}, described explicitly with disordered cells containing thousands symmetry-inequivalent atoms).
First, we have confirmed the inadequacy of the particle-like LBTE in addressing such a problem, as it would predict a conductivity converging to zero upon increasing the size of the disordered cell\cite{garg_thermal_2011}. Second, and most importantly, we have shown that the Wigner transport equation successfully addresses this problem, 
as it describes how in alloys the total conductivity, obtained as sum of the particle-like and wave-like conductivities, converges to the size-independent bulk limit of the conductivity.
The computational scheme introduced here sets the stage to rationalize thermal transport with quantum accuracy in solids with compositional-mass disorder, and will be potentially very useful to develop novel design strategies for thermal barrier coatings. 

\section{Acknowledgements} 
\label{sec:acknowledgements}
M. S. acknowledges support from Gonville and Caius College, and from the SNSF project P500PT\_203178.  A. P., L. B. and N. M. acknowledge support from the Deutsche Forschungsgemeinschaft (DFG) under Germany’s Excellence Strategy (EXC 2077, No. 390741603, University Allowance, University of Bremen) and Lucio Colombi Ciacchi, the host of the “U Bremen Excellence Chair Program”. We also thank the HLRN resource allocation board for granting the computational resources on the supercomputer Lise and Emmy at NHR@ZIB and NHR@G\"{o}ttingen as part of the NHR infrastructure (projects ID:hbp00075 and ID:hbi00059).

\appendix\label{Appendix}
\section{Phonon linewidths} 
\label{sec:contributions_to_phonon_linewidths}

The linewidths appearing in Eq.~(\ref{eq:thermal_conductivity_combined}) are determined by third-order anharmonicity  \cite{paulatto_anharmonic_2013,PhysRevB.88.045430}, isotopic-mass disorder~\cite{PhysRevB.27.858,PhysRevB.88.045430}, and grain-boundary scattering  \cite{fugallo_thermal_2014,casimir_note_1938}.
Specifically, the anharmonic linewidth is:
\begin{equation}
\begin{split}
&\hbar\Gamma^{\rm anh}(\bm{q})_s=\frac{\pi}{\hbar N_c}
{\sum_{\bm{q'},s',s''}}\left 
| V^{(3)}_{\bm{q}s, \bm{q'}s', \bm{q''}s''}\right|^2 \\&
\times \Big\{  \big[1{+}\bar{\tenscomp{N}}(\bm{q'})_{s'}+\bar{\tenscomp{N}}(\bm{q''})_{s''}\big]\delta\big[\omega(\bm{q})_{s}{-}\omega(\bm{q'})_{s'}{-}\omega(\bm{q''})_{s''}\big]\\
&+2\big[\bar{\tenscomp{N}}(\bm{q'})_{s'}{-}\bar{\tenscomp{N}}(\bm{q''})_{s''}\big]\delta\big[\omega(\bm{q})_s+\omega(\bm{q'})_{s'}-\omega(\bm{q''})_{s''}\big]\Big\},
	\label{eq:anh_linewidth}
\end{split}
\raisetag{23mm}
\end{equation}
where 
\begin{equation} 
\begin{split}
V^{(3)}_{\bm{q}s, \bm{q'}s', \bm{q''}s''}=\sum_{\substack{\alpha,\alpha',\alpha''\\b,b',b''}}&
\mathcal{E}(\bm{q})_{s,b\alpha}
\mathcal{E}(\bm{q'})_{s',b'\alpha'}
\mathcal{E}(\bm{q''})_{s'',b''\alpha''} \\
&\sqrt{\frac{1}{f_x(M_b)f_x(M_{b'})f_x(M_{b''})}}\\
&\sqrt{\frac{\hbar^{3}}{8}}\sqrt{\frac{1}{\omega(\bm{q})_s\omega(\bm{q'})_{s'}\omega(\bm{q''})_{s''} }}\\
&\frac{1}{N_c}\frac{\partial^3 E^{tot}}{\partial u(\bm{q})_{b\alpha}\partial u(\bm{q'})_{b'\alpha'}\partial u(\bm{q''})_{b''\alpha''}}	
\end{split}
\label{eq:third_order_matrix_element}
\end{equation}
 are the three-phonon coupling matrix elements  \cite{paulatto_anharmonic_2013} and $f_x(M_b)$ is the mass-replacement function used to account for compositional disorder in the mass-substitution approximation discussed in Sec.~\ref{sub:mass_sub}.
The linewidth due to isotopic-mass disorder (used only in the pure cases) is  \cite{PhysRevB.27.858,PhysRevB.88.045430}
\begin{equation}
\begin{split}
\hbar\Gamma^{\rm iso}(\bm{q})_s=&\frac{\hbar\pi}{2 N_c}[\omega(\bm{q})_s]^2
{\sum_{\bm{q'},s'}} 
\delta\big[\omega(\bm{q})_{s}{-}\omega(\bm{q'})_{s'}\big]\\
&\times\sum_b g_2^b \Big|\sum_\alpha \mathcal{E}(\bm{q})^{\star}_{s,b\alpha} \mathcal{E}(\bm{q'})_{s',b\alpha} \Big|^2,
	\label{eq:linewidts_iso}
\end{split}
\end{equation}
where $g_2^b=\sum_i f_{i,b}\big(\frac{\langle m_b\rangle -m_{i,b}}{\langle m_b\rangle}\big)^2$ is the mass-variance parameter for the masses of atom $b$ ($f_{i,b}$ and $m_{i,b}$ are the mole fraction and mass, respectively, of the $i$th isotope of atom $b$; $\langle m_b\rangle=\sum_i f_{i,b} m_{i,b}$ is the weighted average mass).

Finally, the linewidth due to grain-boundary scattering evaluated according to the Casimir model  \cite{casimir_note_1938} in the presence of perfectly absorbing boundaries is  \cite{fugallo_thermal_2014},
\begin{equation}
	\hbar\Gamma(\bm{q})^{\rm bnd}_s=\frac{\left\Vert \tens{v}(\bm{q})_{ss}\right\Vert}{L}.
	\label{eq:lw_bnd}
\end{equation}
	
\section{Representing anharmonic linewidths as a function of frequency} 
\label{sec:accounting_for_anharmonicity_at_a_reduced_computational_cost}

\begin{figure}[b]
\includegraphics[width=\WidthFigure]{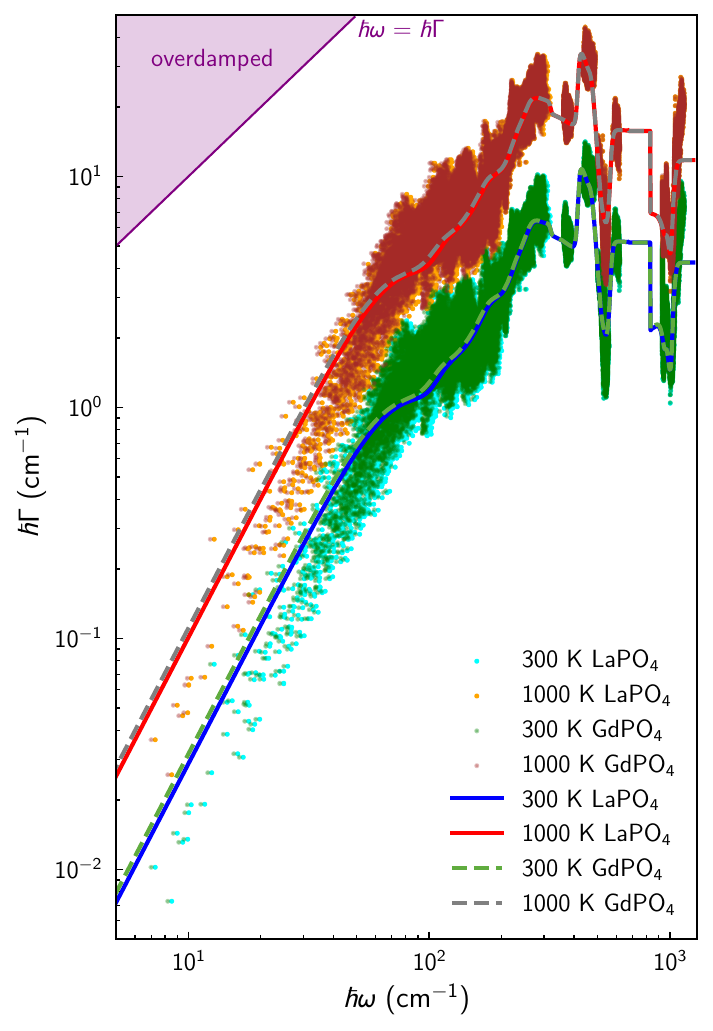}
\caption {\textbf{Frequency-anharmonic linewidth distribution of LaPO$_4$ and GdPO$_4$ at different temperatures,} represented as scatter points. The lines show their coarse-graining into single-valued functions $\Gamma^{\rm a}(\omega)$; coarse-grained functions for LaPO$_4$ are solid, and for GdPO$_4$ are dashed. The purple area is the overdamped regime, where the Wigner formulation cannot be applied. 
} \label{LW_fit-TC-compare}
\end{figure}

In this section, we discuss the details of the computation of the analytical function $\Gamma^{\rm a}(\omega)$, used in Sec.~\ref{sub:mass_sub} to approximatively determine the linewidths as a function of frequency.
Similarly to previous work  \cite{simoncelli_thermal_2023,thebaud_perturbation_2022,garg_thermal_2011}, the description of anharmonic linewidths as a single-value function of frequency, $\Gamma^{\rm a}(\omega)$, is determined as
\begin{equation}
  \Gamma^{\rm a}(\omega)=\frac{1}{\sqrt{\frac{1}{(\Gamma_1(\omega))^2}+\frac{1}{(\Gamma_2(\omega))^2}}},
  \label{eq:approx_analytical_f}
\end{equation}
where $\Gamma_1(\omega)$ and $\Gamma_2(\omega)$ are defined as
\begin{equation}
\Gamma_1(\omega){=}\frac{\sum\limits_{\bm{q},s}\frac{1}{\sqrt{2\pi\sigma^2}}\exp\Big[-\frac{\hbar^2(\omega(\bm{q})_s-\omega)^2}{2\sigma^2}\Big]}{\sum\limits_{\bm{q},s}\tau(\bm{q})_s\frac{1}{\sqrt{2\pi\sigma^2}}\exp\Big[-\frac{\hbar^2(\omega(\bm{q})_s-\omega)^2}{2\sigma^2}\Big]},
\label{eq:approx_analytical_f1}
\end{equation}
\begin{equation}
\begin{split}
    &\Gamma_2(\omega){=} p\cdot \omega^2,\\
    &p=\frac{\sum\limits_{\bm{q},s}\int_{\omega_{\rm o}}^{2\omega_{\rm o}}d\omega' \frac{\Gamma(\bm{q})_s}{\omega^2(\bm{q})_s}
    \frac{2.35}{\sqrt{2\pi\sigma^2}}\exp\!\Big[{-}\tfrac{\hbar^2(\omega(\bm{q})_s-\omega')^2}{2\sigma^2}\Big] }{\sum\limits_{\bm{q},s}\int_{\omega_{\rm o}}^{2\omega_{\rm o}}d\omega'
    \frac{1}{\sqrt{2\pi\sigma^2}}\exp\!\Big[{-}\tfrac{\hbar^2(\omega(\bm{q})_s-\omega')^2}{2\sigma^2}\Big]}.
\end{split}
\label{eq:approx_analytical_f2}
\end{equation}
\begin{figure}[b]
	\centering
	\includegraphics[width=\columnwidth]{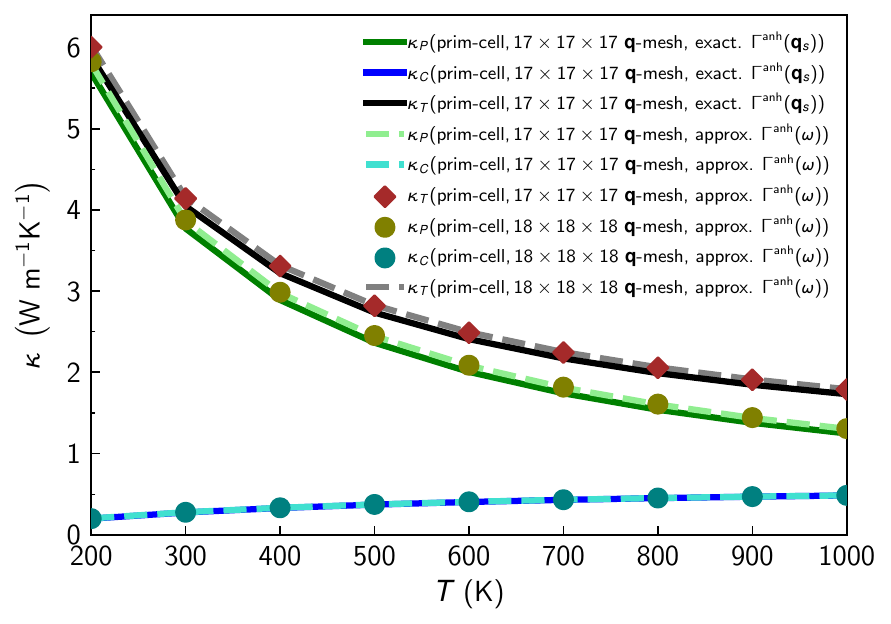}\\[-3mm]
	\caption{\textbf{Validation of the approximated treatment of anharmonicity.}
	In LaPO$_4$ the $\kappa_P$, $\kappa_C$, and $\kappa_T$ conductivities computed using the primitive cell and a $17{\times}17{\times}17$ $\bm{q}$ mesh remain practically unchanged if the exact linewidths (solid) or approximated linewidths (dashed, see Eq.~\ref{eq:approx_analytical_f}) are used.
	 Using the primitive cell and a $18{\times}18{\times}18$ $\bm{q}$ mesh, which contains exactly the same frequencies of a perfect $6{\times}6{\times}6$ supercell with a $3{\times}3{\times}3$ $\bm{q}$ mesh (see Fig.~\ref{fig:LBTE_size_cons}), yields conductivities (scatter points) compatible with the exact ones.
	}
	\label{fig:validation_approx}
\end{figure}
$\omega_{\rm o}$ is the smallest non-zero frequency, and $\sigma{=}15$ cm$^{-1}$ is a broadening chosen sufficiently large to ensure that the linewidths are averaged in a smooth way. The functional form of the approximated function $\Gamma^{\rm a}(\omega)$ is inspired by past work  \cite{garg_thermal_2011,thebaud_perturbation_2022,simoncelli_thermal_2023} and the specific expressions~(\ref{eq:approx_analytical_f},\ref{eq:approx_analytical_f1},\ref{eq:approx_analytical_f2}) to determine it have been devised and validated relying on exact calculations performed in pure LaPO$_4$. Specifically, we show in Fig.~\ref{LW_fit-TC-compare} the function $\Gamma_{\rm LaPO_4}(\omega)$  for pure  LaPO$_4$. In Fig.~\ref{fig:validation_approx} we demonstrate that the exact and approximated treatment of anharmonicity yields practically indistinguishable conductivities over the entire temperature range analyzed.


\section{Single-vibration contribution to the particle-like and wave-like conductivities}  
\label{kp_and_kc}
In this section we report the full expression for the contribution of a vibration with wavevector $\bm{q}$ and mode $s$ to the average trace of the particle-like and wave-like conductivity tensors ($\bar{\mathcal{K}}_{P}(\bm{q})_{s}$ and $\bar{\mathcal{K}}_{C}(\bm{q})_{s}$, respectively). 
These single-mode conductivity contributions are used to compute the size of the points in Fig~\ref{Lifetime-MFP-plot} (size $\propto\bar{\mathcal{K}}_{P}(\bm{q})_{s}{+}\bar{\mathcal{K}}_{C}(\bm{q})_{s}$) and their color~(Eq.~\ref{eq3}). 
We note that this analysis could be extended to resolve Cartesian-direction-dependent contributions to the conductivity, here we consider the average over the three Cartesian directions for simplicity.

The expression for $\bar{\mathcal{K}}_{P}(\bm{q})_{s}$ follows directly from the average trace of the integrand of the SMA particle-like conductivity \cite{PhysRevLett.106.045901}, 
\begin{equation}
\begin{split}
    &\kappa^{av}_{\scriptscriptstyle{\rm P,SMA}} {=}\frac{1}{3}\sum_{\alpha=1}^{3}\frac{1}{\mathcal{V}N_c}\!
\sum\limits_{\bm{q}s}C(\bm{q})_s{\tenscomp{v}^\alpha\hspace*{-0.5mm}(\bm{q})}_{\hspace*{-0.5mm}s,s}\hspace*{-0.4mm}{\tenscomp{v}}^\alpha\hspace*{-0.5mm}(\bm{q})_{\hspace*{-0.5mm}s,s}
  \frac{1}{\Gamma(\bm{q})_{s}},\hspace*{10mm}
\end{split}
\raisetag{8mm}
  \label{eq:SMA_k}
\end{equation}
where $\mathcal{V}$ is the primitive-cell volume, $N_c$ is the number of $\bm{q}$ points appearing in the sum, $C(\bm{q})_s$, ${\tenscomp{v}}^\alpha\hspace*{-0.5mm}(\bm{q})_{\hspace*{-0.5mm}s,s}$ and ${\Gamma(\bm{q})_{s}}$ are the specific heat, group velocity, and linewidths of the phonon $(\bm{q})_s$ already discussed in Sec.~\ref{sec:Wigner_formulation}.
The contributions to the particle-like conductivity  $\bar{\mathcal{K}}_{P}(\bm{q})_{s}$ are obtained from the terms entering in the sum of Eq.~(\ref{eq:SMA_k}):
\begin{equation}
\bar{\mathcal{K}}_{P}(\bm{q})_{s} = C(\bm{q})_s \bar{V}^{\rm av}(\bm{q})_{s,s}\bar{V}^{\rm av}(\bm{q})_{s,s} \left [ \Gamma(\bm{q})_s \right ]^{-1}, \label{eq:kp-av}
\end{equation}
where $ {\bar{V}^{av} \bm (q)_{s,s}}$ = $\sqrt{\frac{1}{3} \sum_{\alpha=1}^{3}\left|\tenscomp{v}^{\alpha} \bm({q})_{s,s} \right|^2 }$ is the spatially averaged group velocity. 

The expression for $\bar{\mathcal{K}}_{C}(\bm{q})_{s}$ is obtained starting from the average trace of the coherences conductivity tensor,
\begin{equation}
\begin{split}
&\kappa^{av}_C{=}\frac{1}{3}\sum_{\alpha=1}^{3}\frac{1}{\mathcal{V}N_c}\sum_{\bm{q},s\neq s'}\frac{\omega(\bm{q})_{s}{+}\omega(\bm{q})_{s'}}{4}\!
\left[\frac{C(\bm{q})_s}{\omega(\bm{q})_s}{+}\frac{C(\bm{q})_{s'}}{\omega(\bm{q})_{s'}}\right]\!\\
&{\tenscomp{v}^\alpha}(\bm{q})_{s,s'\!}{\tenscomp{v}}^\alpha(\bm{q})_{s'\!,s}\frac{\tfrac{1}{2}\big[\Gamma(\bm{q})_{s}{+}\Gamma(\bm{q})_{s'}\big]}{[\omega(\bm{q})_{s'}{-}\omega(\bm{q})_{s}]^2+\tfrac{1}{4}[\Gamma(\bm{q})_{s}{+}\Gamma(\bm{q})_{s'}]^2}.
\end{split}
\label{eq:thermal_conductivity_final_sum}
\end{equation}
The integrand of this equation contains couplings between pairs of vibrational modes. To resolve how much each single vibrational mode contributes to the pairwise wave-like (coherences) transport mechanism, we weight the contribution of each mode with its specific heat (see Ref.~\cite{Wigner_paper} for details), yielding the single-vibration contribution to the wave-like conductivity:
  \begin{equation}
  \begin{split}
&\bar{\mathcal{K}}_C(\bm{q})_s{=} \!\!
\sum_{s'\neq s}\!\!
  \frac{C(\bm{q})_s}{C(\bm{q})_{s}{+}C(\bm{q})_{s'\!}\!}
\frac{\omega(\bm{q})_{s}{+}\omega(\bm{q})_{s'}\!}{2}\!\!
\left[\!\frac{C(\bm{q})_s}{\omega(\bm{q})_s}{+}\frac{C(\bm{q})_{s'\!}}{\omega(\bm{q})_{s'\!\!}}\!\right]\!\\
&{\times}\!\!
\left[\frac{1}{3}\!\sum_\alpha|{\tenscomp{v}^\alpha}(\bm{q})_{s,s'\!}|^2\!\right]\!\!\frac{\tfrac{1}{2}\big[\Gamma(\bm{q})_{s}{+}\Gamma(\bm{q})_{s'}\big]}{[\omega(\bm{q})_{s'}{-}\omega(\bm{q})_{s}]^2{+}\tfrac{1}{4}[\Gamma(\bm{q})_{s}{+}\Gamma(\bm{q})_{s'}]^2}.
  \end{split}
\label{eq:coherence_density}
  \end{equation}

\begin{table*}
\centering{}%
\begin{tabular}{>{\centering}p{5cm}>{\centering}p{1.5cm}>{\centering}p{1.5cm}>{\centering}p{1.5cm}>{\centering}p{1.5cm}>{\centering}p{1.5cm}>{\centering}p{1.5cm}>{\centering}p{1.5cm}}
\toprule 
Functional & \textit{\small{}a $(\lyxmathsym{\AA})$} & \textit{\small{}b $(\lyxmathsym{\AA})$} & \textit{\small{}c $(\lyxmathsym{\AA})$} & {\small{}$\alpha(\lyxmathsym{\textdegree})$} & {\small{}$\beta(\lyxmathsym{\textdegree})$} & {\small{}$\gamma(\lyxmathsym{\textdegree})$} & \textit{\small{}V $(\lyxmathsym{\AA}^{3})$}\tabularnewline
\midrule
\addlinespace
{\small{}PBEsol} & {\small{}6.843} & {\small{}7.074} & {\small{}6.478} & {\small{}90.000} & {\small{}103.478} & {\small{}90.000} & {\small{}304.938}\tabularnewline
\addlinespace
{\small{}Exp. from Ni et al.~\cite{Ni-et.al}} & {\small{}6.831} & {\small{}7.071} & {\small{}6.503} & {\small{}90.000} & {\small{}103.270} & {\small{}90.000} & {\small{}305.732}\tabularnewline
\bottomrule
\addlinespace
\end{tabular}\caption{\label{tab:Optimized-Lattice}Optimized lattice parameters and volume of LaPO$_{4}$. Experimental data are taken from Ni et al.~\cite{Ni-et.al}. The theoretical primitive-cell volume is 0.26 $\%$ smaller than the experimental measure at 300 K.}
\end{table*}

\section{Strength of particle-like \& wave-like transport in ordered or disordered cells } 
\label{sec:size_consistency_of_the_lbte_in_pristine_systems}
\begin{figure}[b]
	\centering
	\includegraphics[width=\columnwidth]{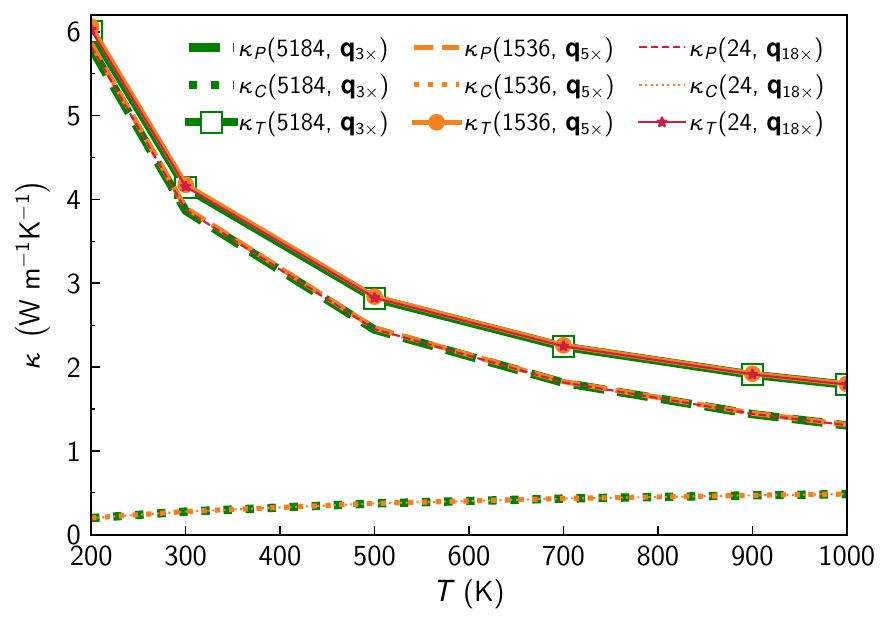}
	\caption{
\textbf{Conductivity of pristine LaPO$_4$ described with different unit cells.}	
The particle-like conductivity ($\kappa_P$, dashed, equivalent to the Peierls-Boltzmann conductivity), wave-like conductivity (dotted, $\kappa_C$), and total Wigner conductivity (solid, $\kappa_T{=}\kappa_P{+}\kappa_C$) are independent from the arbitrary choice of the unit cell used to describe a pristine crystal. Using the 24-atom primitive cell (red, evaluated on a $18{\times}18{\times}18$ $\bm{q}$ mesh), or supercells of the primitive containing a number of atoms equal to 1536 (orange, $4{\times}4{\times}4$ supercell, evaluated on a $5{\times}5{\times}5$ $\bm{q}$ mesh) and 5184 (green, $6{\times}6{\times}6$ supercell, evaluated on a $3{\times}3{\times}3$ $\bm{q}$ mesh), yields indistinguishable results.}
	\label{fig:LBTE_size_cons}
\end{figure}

We show in Fig.~\ref{fig:LBTE_size_cons} that in pristine LaPO$_4$, the particle-like, wave-like, and total conductivities do not depend on the arbitrary choice of the unit cell used to describe the system, in agreement with physical expectations.

We now provide an analytical reasoning that complements the numerical results reported in Sec.~\ref{ssub:comp_dis_vib_prop}, which showed that in the presence of disorder the particle-like conductivity decreases upon increasing the size of the disordered cell used to describe the system (Figs.~\ref{supecercell-convergence},\ref{fig:alloy_k}).
This behavior can be understood analytically, recalling that: (i) the conductivity of an extremely large disordered cell can be described without relying on Fourier interpolation (\textit{i.e.}, evaluating Eq.~\ref{eq:thermal_conductivity_combined} at $\bm{q}=\bm{0}$ only, or over a $n{\times}n{\times}n$ $\bm{q}$-mesh would produce no differences, see Ref.~\cite{simoncelli_thermal_2023} for details); (ii) the presence of disorder forbids degeneracies between vibrational modes\cite{maradudin_symmetry_1968}; (iii) due to time-reversal symmetry, the velocity operator at $\bm{q}{=}\bm{0}$ has zero diagonal elements. These properties imply that in the limiting case of an infinitely large disordered cell $\kappa_P=0$, and only $\kappa_C$ determines the total conductivity. 
To further confirm the correctness of this analytical reasoning, it is useful to apply it to the opposite limit of a perfect supercell of a pristine crystal. In this case, the vibrational frequencies at $\bm{q}{=}\bm{0}$ of the perfect supercell are simply obtained folding the phonon bands of the primitive cell \cite{mayo_band_2020}. Then, the presence of crystal symmetries within the supercell allows for the emergence of perfectly degenerate vibrational modes, which contribute to the particle-like conductivity through non-zero off-diagonal-and-degenerate velocity-operator elements.
These considerations imply that in pristine crystals $\kappa_P$, $\kappa_C$, and $\kappa_T$ do not depend on the arbitrary choice of the unit cell used to describe the conductivity (see dashed lines in Fig.~\ref{fig:LBTE_size_cons} and Fig.~\ref{fig:alloy_k}, as well as Fig. 2 of Ref.~\cite{Wigner_paper}).

\section{Negligible effect of isotopes on the conductivity} 
\label{sec:negligible_effect_of_isotopes_on_the_conductivity}
\begin{figure}[b]
\centering
\includegraphics[width=\columnwidth]{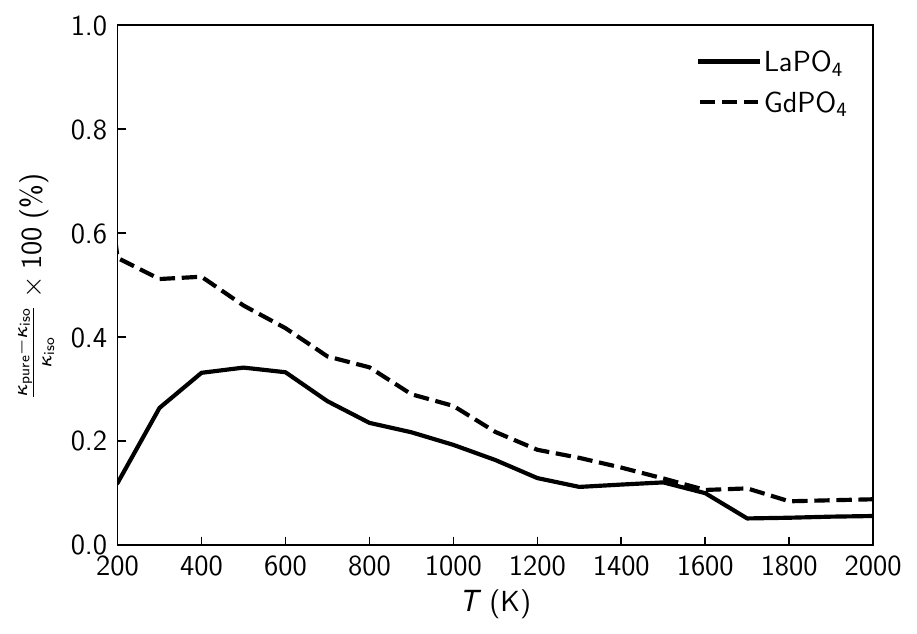}
\caption{\textbf{Negligible effect of isotopes on conductivity.} 
In pristine LaPO$_4$ (solid) and GdPO$_4$ (dashed), the relative difference between the total conductivity obtained accounting or not for isotopes at natural abundance\cite{PhysRevB.27.858} ($\kappa_{\rm iso}(T)$ and $\kappa_{\rm pure}(T)$, respectively) is always smaller than $0.6\%$.
}
\label{fig:negligible_ISO}
\end{figure}
In Fig.~\ref{fig:negligible_ISO} we show that accounting or not for the contributions of isotopes to the linewidths\cite{PhysRevB.27.858} (Eq.~(\ref{eq:linewidts_iso})) has negligible effects on the thermal conductivities of LaPO$_4$ and GdPO$_4$ (variations smaller than $1\%$).

\section{Computational methods}
\label{sec: computational methods}
\subsection{Structural and vibrational properties of LaPO$_{4}$} 
\label{sub:structural_and_vibrational_properties}

\begin{figure}[b] 
\begin{centering}
\includegraphics[width=0.4\columnwidth]{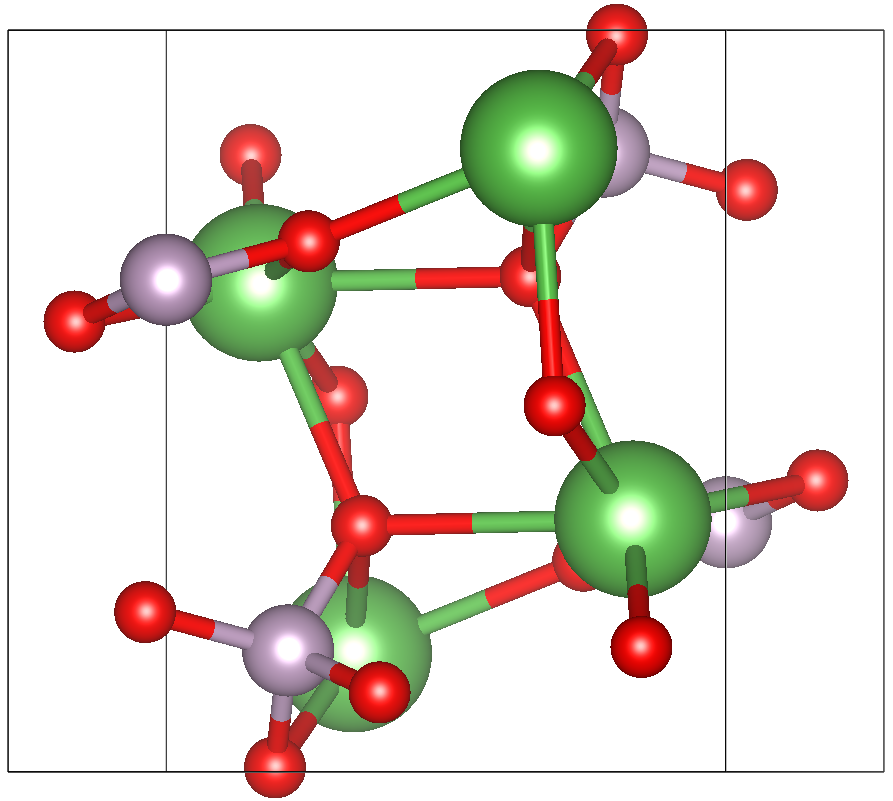}\caption{\textbf{DFT-optimized primitive cell of LaPO$_{4}$.} Green, purple, and red represent La, P, and O atoms, respectively.}\label{optimized-str}
\par\end{centering}
\end{figure}

The crystal structure of monazite LaPO$_{4}$ is taken from the experimental work of Ni et al.~\cite{Ni-et.al}, it is monoclinic (space group $P2_{1}/n$) and contains 24 atoms (4 formula units) in the primitive cell (Fig. \ref{optimized-str}).  

DFT calculations are performed using the Quantum ESPRESSO (QE) distributions \cite{Giannozzi_2017}. 
We employed the revised Perdew-Burke-Enzerhof functional (PBEsol) \cite{PhysRevLett.100.136406}, pseudo-potentials were taken from the SSSP precision library (version 1.1.2)  \cite{prandini_precision_2018,lejaeghere_reproducibility_2016}. 
We used a kinetic energy cut-off of 80 Ry, and the Brillouin zone was sampled using a $3{\times}3{\times}3$
Monkhorst-Pack  \cite{PhysRevB.13.5188} k-point mesh with a (1 1 1) shift. 
The structure is relaxed using the variable cell relax (vc-relax) scheme,  with a force convergence threshold of 10$^{-5}$ Ry/Bohr. 
The resulting equilibrium lattice parameters are in good agreement with experiments, see Table \ref{tab:Optimized-Lattice}. 

\begin{figure}[h] 
\begin{centering}
\includegraphics[width=\WidthFigure]{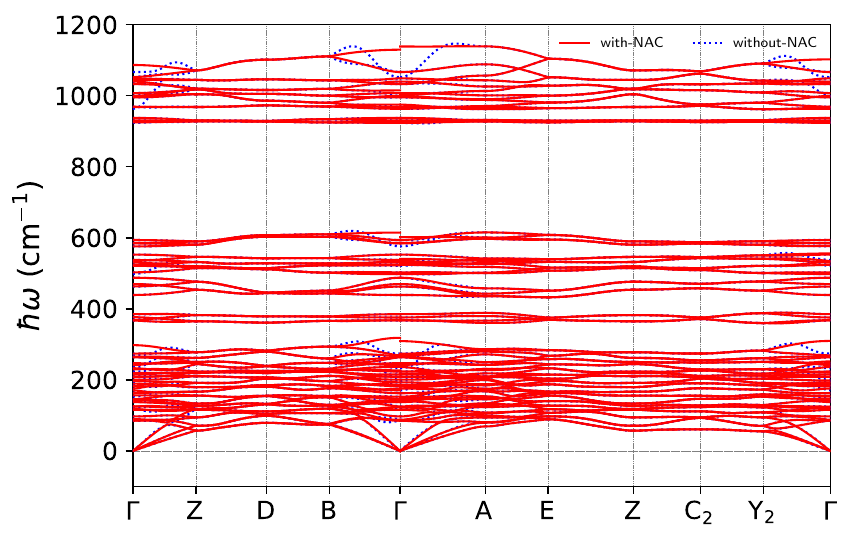}\caption{\textbf{Phonon dispersion of LaPO$_{4}$.}  
Red (blue) is with (without) non-analytic (NAC) term correction. We highlight how NAC affects the splitting of LO-TO mode at $\bm{q}=\bm{0}$. }\label{phonon-disp}
\par\end{centering}
\end{figure}

The second-order force constants (fc2) are computed using density functional perturbation theory (DFPT) \cite{RevModPhys.73.515} over a $4{\times}4{\times}4$ $\bm{q}$ mesh in reciprocal space.  The LO-TO splitting at the $\Gamma$ point is incorporated using the non-analytic term correction computed with dielectric tensor and Born effective charges \cite{Gonze1997}. The absence of the imaginary phonon modes in the phonon dispersion (Fig. \ref{phonon-disp}) confirms the dynamical stability. 

The third-order force constants (fc3) are computed with a $ 2\times2\times2\ $  supercell using QE and the ShengBTE packages  \cite{LI20141747}.  Here, the finite differences method is used and the nearest neighbor (nn) interaction up to 8 nn is incorporated. The QE fc2 and fc3 are exported to hdf5 formats using Phonopy  \cite{TOGO20151} and HIPHIVE   \cite{https://doi.org/10.1002/adts.201800184} packages, respectively. The linewidths are then computed using the Phono3py package  \cite{PhysRevB.91.094306,togo_first-principles_2023}.

\begin{figure*}
\includegraphics[scale=0.65]{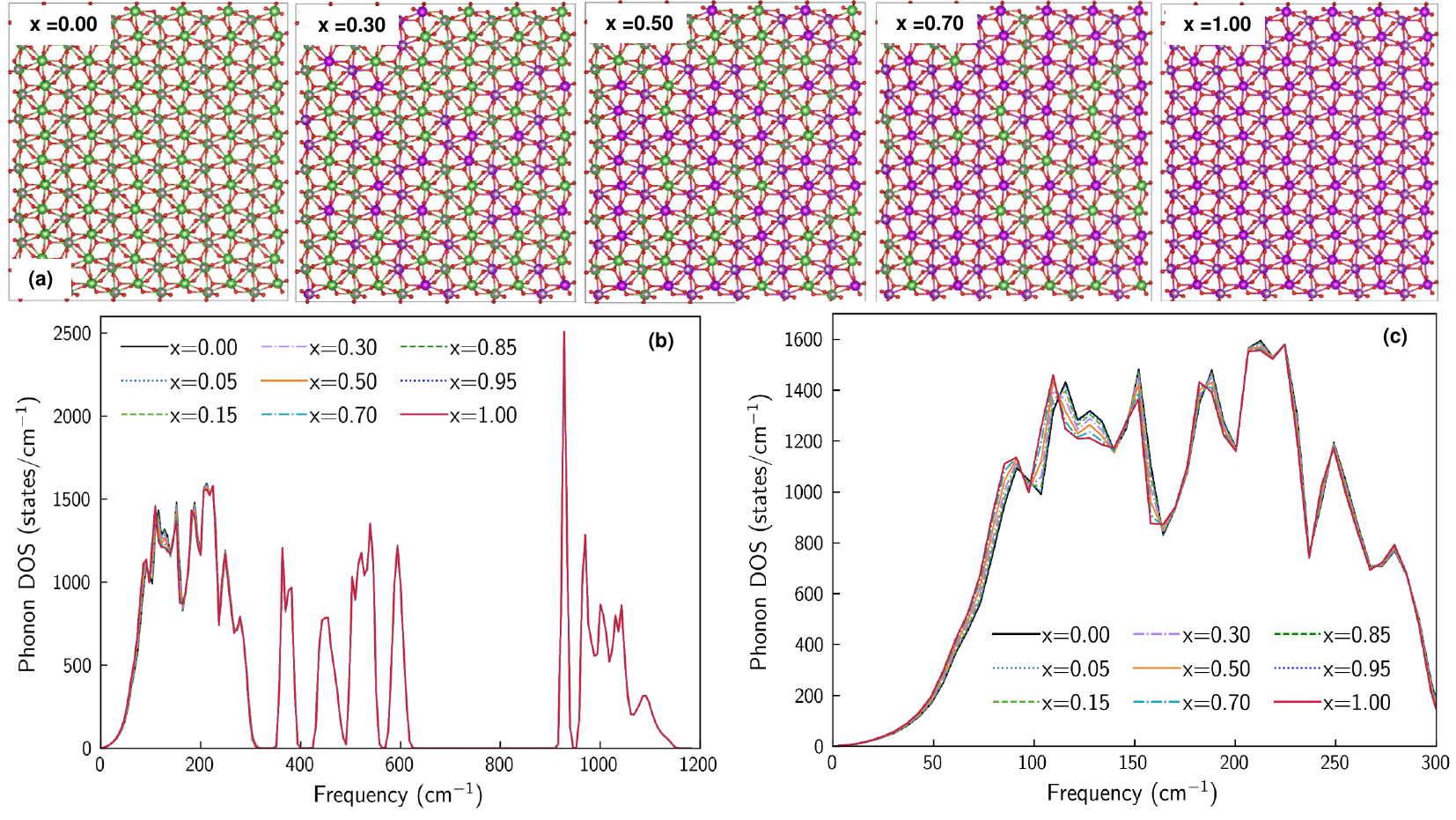}\\[-3mm]
\caption{\textbf{Explicit simulation of harmonic vibrational properties of La$_{1-x}$Gd$_{x}$PO$_{4}$ alloys.} Panel a), front view of the  5184-atom disordered cell of La$_{1-x}$Gd$_{x}$PO$_{4}$ of different compositions $x$. The green, violet, red, and purple colors represent La, P, O, and Gd atoms, respectively. Panel b), phonon density of states as a function of $x$; for all the disordered configurations considered, the vibrational frequencies are real and positive, indicating dynamic stability. Panel c), zoom of phonon DOS in the range 50 to 300 cm$^{-1}$, where compositional disorder has the largest effects.}\label{Alloy-config-DOS}
\end{figure*}

\subsection{Raman spectrum} 
\label{sub:raman_spectrum}
The Raman intensities $I_s$ appearing in Eq.~(\ref{eq:raman}) were computed from the Raman tensor\cite{Bastonero2024,Umari2005},
\begin{equation}
    \frac{\partial \chi_{ij} }{\partial u_{k,I} }
    =
    \frac{1}{\Omega}
    \frac{\partial^2 F_{k,I}}{\partial \mathcal{E}_i \partial \mathcal{E}_j}
    \label{eq:RamanTensor}
\end{equation}
where  $F_{k,I}$ is the force acting on atom $I$ and $\mathcal{E}$ is the macroscopic electric field. 
The Raman tensor was computed using the finite-electric-field approach \cite{Umari2005} as implemented in the \texttt{aiida-vibroscopy} package \cite{Bastonero2024} within the AiiDA infrastructure \cite{Huber2020,Uhrin2021}. 
The second-order derivative appearing in Eq.~(\ref{eq:RamanTensor}) was evaluated through the application of a small electric field, described by the electric-enthalpy functional  \cite{Souza2002,Umari2002}, an extension of the Kohn-Sham functional that allows us to find meta-stable solutions in the presence of a homogeneous electric field. In particular, we used a fourth-order central difference formula with an electric field step of about $0.8 \times 10^{-3} ~ \text{(Ry) a.u.}$ (1 (Ry) a.u. $\approx 36.3609$ V/$\AA$) to remove the finite size dependence of the numerical derivative, see Ref.~\cite{Bastonero2024} for details, and a  Monkhorst-Pack grid of $10\times9\times10$ to ensure a well-converged spectra.
Finally, the tensor was symmetrized according to the LaPO$_4$ space group. 

We conclude this section discussing in more detail the temperature dependence of the Raman spectra in Fig.~\ref{fig:raman}. Specifically, we note that the relative intensity of the theoretical Raman spectra does not decrease monotonically with temperature for all the peaks --- the peak around 200 cm$^{-1}$ is more intense at 1000 K than at 300 K. Such a behavior originates from the presence of the temperature-independent instrumental linewidth in the Lorentzian~\ref{eq:raman}, and from the Bose-Einstein distribution appearing in the  Raman cross section for the $I_s$.
To quantitatively understand this behavior, we consider the maximum Raman intensity at $\omega \approx$ 200 {\rm cm}$^{-1}$,
\begin{equation}
    I(\omega=\omega_s,T) \sim (\bar{\tenscomp{N}}(\omega_s,T)+1)/(\Gamma_s + \Gamma_{\rm ins}), 
    \label{eq:intensitiescomparison}
\end{equation}
where $(\bar{\tenscomp{N}}(\omega_s,T)+1) = \left ( 1 - \exp \{-\hbar \omega_s / k_B T \} \right )^{-1}$.
We see from Fig.~\ref{LW_fit-TC-compare} that for $\omega_s \approx$ 200 {\rm cm}$^{-1}$  we have 2 \rm cm$^{-1}=\Gamma_{\rm ins}\gg \Gamma_s $. It follows that the temperature dependence of the Raman intensity of these modes arises entirely from the Bose-Einstein occupations. In particular, comparing the intensities at $T_1$ = 300 K and $T_2 $= 1000 K, we have
\begin{equation}
    I(\omega_s, T_2)/I(\omega_s, T_1)
    \approx 
    (\bar{\tenscomp{N}}(\omega_s,T_2)+1)/(\bar{\tenscomp{N}}(\omega_s,T_1)+1)
    \approx 2.5 
    ~,
\end{equation}
and this explains why the Raman peak at low-frequency becomes sharper increasing temperature.
In contrast, for high-frequency modes, 
2 $\rm cm^{-1}=\Gamma_{\rm ins}\ll \Gamma_s \sim T$, and from Eq.~(\ref{eq:intensitiescomparison}) it follows that
$(\bar{\tenscomp{N}}(\omega_s,T)+1)/\Gamma_s$ decreases upon increasing temperature.  

Finally, we note that the results reported in Fig.~\ref{fig:raman} are computed using the same level of theory used in the thermal conductivity calculations. Specifically, they are computed using the standard perturbative treatment of anharmonicity (used also in the phonon Boltzmann transport equation\cite{peierls_quantum_2001,ziman_electrons_2001}) that considers third-order broadening effects (bubble three-phonon diagram) and neglects the renormalization of the frequencies due to anharmonicity and temperature.
As discussed in Sec.~V of Ref.~\cite{Wigner_paper}, rigorously accounting for the influence of these effects on the thermal conductivity is an open challenging problem, which requires accounting also for anharmonic terms in the heat flux~\cite{hardy_energy-flux_1963} to ensure a consistent treatment of the approximations that have to be performed.

\subsection{Evaluating the Wigner conductivity of alloys} 
\label{sub:thermal_conductivity}
The thermal conductivity is calculated by solving the linearized form of the Wigner transport equation (LWTE) in the homogeneous regime, relying on the solver implemented in the Phono3py code  \cite{PhysRevB.91.094306}. The scattering operator is computed on a mesh of size $ 17\times17\times17\ $ by accounting for the isotopic scattering effects  \cite{PhysRevLett.106.045901,PhysRevB.27.858} and third-order anharmonicity.

\begin{figure}[H]
\centering
\includegraphics[width=\WidthFigure]{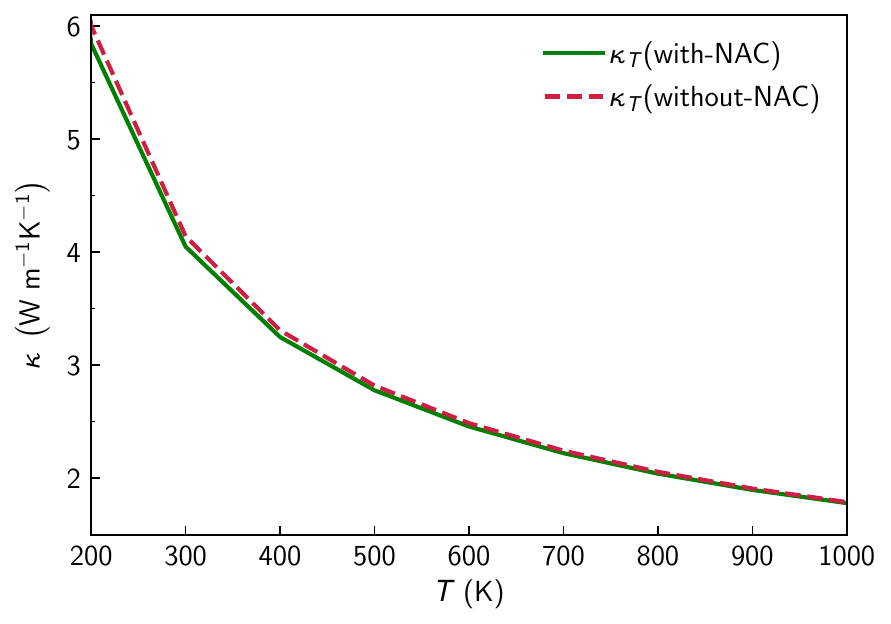}\\[-3mm]
\caption{\textbf{Negligible effect of non-analytical term correction on the conductivity of pristine LaPO${_4}$}, computed using a perfect 5184-atom supercell (a $6{\times}6{\times}6$ repetition of the primitive cell) on a $3{\times}3{\times}3$ $\bm{q}$ mesh.}\label{Effect-BORN}
\end{figure}

To evaluate the effect of compositional disorder on the conductivity of in La$_{(1-x)}$Gd$_{x}$PO$_{4}$, we employ the following computational protocol:
\begin{enumerate}
	\item We  build a perfect $6\times6\times6$ supercell of the primitive cell of LaPO$_4$.
	\item We construct second-order force constants (fc2) for the perfect supercell. We do this by mapping the primitive-cell force-constants into the supercell using the tool of Ref.~\cite{Berges2017}. 
	\item We use the pristine-supercell fc2 computed at the previous point as starting point to describe compositional disorder at 8 different compositions $x\in$[0.05, 0.15, 0.30, 0.50, 0.70, 0.85, 0.95, 1.00]. 
	Specifically, we use the mass-substitution approximation explained in Sec.~\ref{sub:mass_sub} (see Eq.~(\ref{eq:matrix_G}) and Eq.~(\ref{eq:fun_repl})). For each composition $x$ we scan the supercell sites containing La, and with probability $x$, we replace the mass of La atoms with the mass of Gd (representative configurations are shown in Fig. \ref{Alloy-config-DOS}\textbf{a}).
	\item We assess the dynamical stability of each of the 8 compositionally disordered configurations by diagonalizing the mass-substituted, disordered dynamical matrix~(\ref{eq:matrix_G}), and verifying that the vibrational density of states is non-zero only for positive values of the vibrational energy (Fig.~\ref{Alloy-config-DOS}\textbf{b},\textbf{c}).
	\item We compute the velocity operator (Eq.~\ref{eq:vel_op}) for each disordered composition on the computationally converged $3\times3\times3$ $\bm{q}$ mesh (see Fig.~\ref{fig:LBTE_size_cons} for convergence tests). 
	In this step, to reduce the computational cost, we neglect the effect of the non-analytical term correction\cite{Gonze1997}. We validate this approximation in Fig.~\ref{Effect-BORN} by showing its negligible effect on the thermal conductivity.
	\item We employ the velocity operator and the coarse-grained linewidth (Eq.~\ref{eq:mix_lw}) to evaluate the Wigner thermal conductivity~(\ref{eq:thermal_conductivity_combined}). 
\end{enumerate}

We note, in passing, that we considered one disordered configuration for each composition, since the convergence tests in Fig.~\ref{supecercell-convergence} show that the total conductivities of different disordered cells of La$_{0.5}$Gd$_{0.5}$PO$_4$ larger than 1536 atoms the same composition $x=0.5$ are practically indistinguishable.\\


\subsection{Data availability} 
The raw data needed to reproduce the findings of this study are available on the Materials Cloud Archive \cite{Materials_cloud_release}.

\bibliographystyle{apsrev4-2}

%

\end{document}